\documentclass[reprint,NumberedRefs]{JASAnew}

\usepackage{natbib}
\usepackage{graphicx,color,rotating}
\usepackage[latin1]{inputenc}
\usepackage{textcomp}
\usepackage{dcolumn}
\usepackage{bm}     
\usepackage{upgreek}
\usepackage{subdepth}  			
\usepackage{epstopdf}   		
\usepackage[latin1]{inputenc}
\usepackage{amsmath}
\usepackage{amssymb}
\usepackage{amsfonts}
\usepackage{array}
\newcolumntype{L}[1]{>{\raggedright\let\newline\\\arraybackslash\hspace{0pt}}m{#1}}
\newcolumntype{C}[1]{>{\centering\let\newline\\\arraybackslash\hspace{0pt}}m{#1}}
\newcolumntype{R}[1]{>{\raggedleft\let\newline\\\arraybackslash\hspace{0pt}}m{#1}}

\usepackage{xcolor}						
\definecolor{darkgreen}{rgb}{0.00, 0.50, 0.00}
\definecolor{DARKGREEN}{rgb}{0.00, 0.50, 0.00}
\definecolor{RED}{rgb}{1.00, 0.00, 0.00}
\definecolor{GREEN}{rgb}{0.00, 1.00, 0.00}
\definecolor{BLUE}{rgb}{0.00, 0.00, 1.00}
\definecolor{MAGENTA}{rgb}{1.00, 0.00, 1.00}

\newcommand{\qmarks}[1]{{``#1''}}

\newcommand{\mr}[1]{\ensuremath{\mathrm{#1}}}
\newcommand{\myvec}[1]{\bm{#1}}
\newcommand{\ee}{\mathrm{e}}
\newcommand{\ii}{\mathrm{i}}
\newcommand{\dm}{\mathrm{d}}

\newcommand{\avr}[1]{\big\langle #1 \big\rangle}
\newcommand{\abs}[1]{\big|{#1}\big|}

\newcommand{\intII}[2]{I_{{#1}{#2}}^{(2)}}
\newcommand{\intIII}[2]{I_{{#1}{#2}}^{(3)}}
\newcommand{\intn}[2]{I_{{#1}{#2}}^{(n)}}

\DeclareMathOperator{\re}{Re}

\newcommand{\iot}{{\ii\omega t}}

\newcommand{\pp}{\partial}
\newcommand{\ppsqr}{\partial^{\,2_{}}}

\newcommand{\nablabf}{\boldsymbol{\nabla}}

\newcommand{\Lapl}{\nabla^2}

\newcommand{\grad}{\nablabf}
\newcommand{\rot}{\nablabf\times}
\newcommand{\divop}{\nablabf\cdot}

\newcommand{\scap}{\!\cdot\!}

\newcommand{\AAA}{\myvec{A}}

\newcommand{\BBB}{\myvec{B}}

\newcommand{\eee}{\myvec{e}}
\newcommand{\een}{\myvec{e}}

\newcommand{\fff}{\myvec{f}}

\newcommand{\fffac}{\fff_\mathrm{ac}}

\newcommand{\III}{\myvec{I}}

\newcommand{\kc}{k_\mathrm{c}}
\newcommand{\kt}{k_\mathrm{t}}
\newcommand{\ks}{k_\mathrm{s}}
\newcommand{\kcsqr}{k^{2_{}}_\mathrm{c}}
\newcommand{\ktsqr}{k^{2_{}}_\mathrm{t}}
\newcommand{\kssqr}{k^{2_{}}_\mathrm{s}}

\newcommand{\nnn}{\myvec{n}}

\newcommand{\rrr}{\myvec{r}}

\newcommand{\sss}{\myvec{s}}

\newcommand{\uuu}{\myvec{u}}

\newcommand{\VVV}{\myvec{V}}

\newcommand{\vvv}{\myvec{v}}

\newcommand{\vIdflz}{v^{d,\mr{fl}}_{1,z}}
\newcommand{\vIdOflz}{v^{d0,\mr{fl}}_{1,z}}

\newcommand{\vIdOslz}{v^{d0,\mr{sl}}_{1,z}}

\newcommand{\vIdelOflz}{v^{\delta0,\mr{fl}}_{1,z}}
\newcommand{\vvvId}{\myvec{v}^d_1}
\newcommand{\vvvIdO}{\myvec{v}^{d0}_1}
\newcommand{\vvvIdfl}{\myvec{v}^{d,\mr{fl}}_1}
\newcommand{\vvvIdOfl}{\myvec{v}^{d0,\mr{fl}}_1}

\newcommand{\vvvIdOsl}{\myvec{v}^{d0,\mr{sl}}_1}
\newcommand{\vvvIdel}{\myvec{v}^\delta_1}
\newcommand{\vvvIdelO}{\myvec{v}^{\delta0}_1}

\newcommand{\vvvIdelOfl}{\myvec{v}^{\delta0,\mr{fl}}_1}

\newcommand{\zerovec}{\boldsymbol{0}}

\newcommand{\calL}{\mathcal{L}}

\newcommand{\cO}{c_0}

\newcommand{\cp}{c_p}
\newcommand{\cpO}{c_{p0}}

\newcommand{\DthO}{D^\mathrm{th}_0}

\newcommand{\Eac}{E_\mathrm{ac}}
\newcommand{\Lac}{\calL_\mathrm{ac}}
\newcommand{\fac}{\myvec{f}_\mathrm{ac}}
\newcommand{\fachat}{\hat{\myvec{f}}_\mathrm{ac}}

\newcommand{\rhoac}{\dot{\rho}_\mathrm{ac}}

\newcommand{\kth}{k^\mathrm{th}}

\newcommand{\kthO}{k^\mathrm{th}_0}
\newcommand{\kthOsl}{k^\mathrm{th,sl}_0}
\newcommand{\kthOfl}{k^\mathrm{th,fl}_0}
\newcommand{\kthsl}{k^\mathrm{th,sl}}
\newcommand{\kthfl}{k^\mathrm{th,fl}}

\newcommand{\kthIsl}{k^\mathrm{th,sl}_1}
\newcommand{\kthIfl}{k^\mathrm{th,fl}_1}
\newcommand{\kthIIsl}{k^\mathrm{th,sl}_2}
\newcommand{\kthIIfl}{k^\mathrm{th,fl}_2}

\newcommand{\kapT}{\kappa_T}
\newcommand{\kapTO}{\kappa_{T0}}
\newcommand{\kapS}{\kappa_s}

\newcommand{\kaps}{\kappa_s}
\newcommand{\kapsO}{\kappa_{s0}}

\newcommand{\alphap}{{\alpha_p}}

\newcommand{\alfP}{{\alpha_p}}
\newcommand{\alfPO}{{\alpha_{p0}}}

\newcommand{\delt}{\delta_\mathrm{t}}

\newcommand{\dels}{\delta_\mathrm{s}}

\newcommand{\etaB}{\eta^\mathrm{b}}
\newcommand{\etaBO}{\eta^\mathrm{b}_0}

\newcommand{\etaO}{\eta_0}

\newcommand{\Gamt}{\Gamma_\mathrm{t}}
\newcommand{\Gams}{\Gamma_\mathrm{s}}
\newcommand{\GamOcfl}{\Gamma^\mr{fl}_\mathrm{0c}}
\newcommand{\GamOtfl}{\Gamma^\mr{fl}_\mathrm{0t}}

\newcommand{\GamOcsl}{\Gamma^\mr{sl}_\mathrm{0c}}
\newcommand{\GamOtsl}{\Gamma^\mr{sl}_\mathrm{0t}}

\newcommand{\cOsqr}{c^{\,2_{}}_0}

\newcommand{\kOsqr}{k^{2_{}}_0}

\newcommand{\pI}{p_1}

\newcommand{\TId}{T^d_1}

\newcommand{\TIdOfl}{T^{d0,\mr{fl}}_1}

\newcommand{\TIdOsl}{T^{d0,\mr{sl}}_1}
\newcommand{\TIdel}{T^\delta_1}

\newcommand{\TIdelfl}{T^{\delta,\mr{fl}}_1}
\newcommand{\TIdelOfl}{T^{\delta0,\mr{fl}}_1}
\newcommand{\TIdelsl}{T^{\delta,\mr{sl}}_1}
\newcommand{\TIdelOsl}{T^{\delta0,\mr{sl}}_1}

\newcommand{\TI}{T_1}

\newcommand{\vvvI}{\vvv_1}

\newcommand{\vvvIsl}{\vvvI^\mr{sl}}

\newcommand{\rhoO}{\rho_0}

\newcommand{\rhoI}{\rho_1}

\newcommand{\SICel}{^\circ\!\textrm{C}}

\newcommand{\SIum}{\upmu\textrm{m}}

\newcommand{\SIGHz}{\textrm{GHz}}
\newcommand{\SIMHz}{\textrm{MHz}}

\newcommand{\SIJ}{\textrm{J}}

\newcommand{\SIK}{\textrm{K}}

\newcommand{\SIm}{\textrm{m}}

\newcommand{\SImm}{\textrm{mm}}
\newcommand{\SImum}{\textrm{\textmu{}m}}

\newcommand{\SInm}{\textrm{nm}}

\newcommand{\SIMPa}{\textrm{MPa}}

\newcommand{\SIs}{\textrm{s}}

\newcommand{\nn}{\nonumber}
\newcommand{\beq}[1]{\begin{equation} \eqlab{#1}}
\newcommand{\eeq}{\end{equation}}
\newcommand{\bsub}{\begin{subequations}}
\newcommand{\esub}{\end{subequations}}
\def\bal#1\eal{\begin{align}#1\end{align}}
\def\balat#1#2\ealat{\begin{alignat}{#1} #2 \end{alignat}}
\def\bsubal#1 #2\esubal{\bsuba{#1}\begin{align}#2\end{align} \esuba}     
\def\bsubalat#1#2#3\esubalat{\bsuba{#1} \begin{alignat}{#2} #3 \end{alignat} \esuba}
\newcommand{\bsuba}[1]{\bsub \eqlab{#1}}
\newcommand{\esuba}{\esub}

\newcommand{\eqlab}[1]{\label{eq:#1}}
\renewcommand{\eqref}[1]{Eq.~(\ref{eq:#1})}
\newcommand{\eqnoref}[1]{(\ref{eq:#1})}

\newcommand{\eqsref}[2]{Eqs.~(\ref{eq:#1}) and~(\ref{eq:#2})}
\newcommand{\eqsnoref}[2]{(\ref{eq:#1}) and~(\ref{eq:#2})}

\newcommand{\figref}[1]{Fig.~\ref{fig:#1}}

\newcommand{\figlab}[1]{\label{fig:#1}}

\newcommand{\secref}[1]{Section~\ref{sec:#1}}
\newcommand{\secnoref}[1]{\ref{sec:#1}}

\newcommand{\seclab}[1]{\label{sec:#1}}

\newcommand{\Gamv}{\Gamma_{\mathrm{s}}}

\newcommand{\vPsi}{\myvec{\Psi}}

\newcommand{\phic}{\phi_\mathrm{c}}
\newcommand{\phit}{\phi_\mathrm{t}}

\newcommand{\sigmabf}{\bm{\sigma}}

\newcommand{\taubf}{\bm{\tau}}

\newcommand{\cL}{c_\mathrm{lo}}
\newcommand{\cT}{c_\mathrm{tr}}

\newcommand{\cLsqr}{c^2_\mathrm{lo}}
\newcommand{\cTsqr}{c^2_\mathrm{tr}}

\newcommand{\uuuI}{\myvec{u}_1}
\newcommand{\uuuIO}{\myvec{u}^0_1}
\newcommand{\uuuIL}{\myvec{u}^\mr{lo}_1}
\newcommand{\uuuIT}{\myvec{u}^\mr{tr}_1}

\begin{document}

\title{Theory of pressure acoustics with thermoviscous boundary layers and streaming in elastic cavities}

\author{Jonas Helboe Joergensen}
\email{jonashj@fysik.dtu.dk}
\affiliation{Department of Physics, Technical University of Denmark,\\ DTU Physics Building 309, DK-2800 Kongens Lyngby, Denmark}

\author{Henrik Bruus}
\email{bruus@fysik.dtu.dk}
\affiliation{Department of Physics, Technical University of Denmark,\\
DTU Physics Building 309, DK-2800 Kongens Lyngby, Denmark}

\date{14 December 2020}

\begin{abstract}

We present an effective thermoviscous theory of acoustofluidics including pressure acoustics, thermoviscous boundary layers, and streaming for fluids embedded in elastic cavities. By including thermal fields, we thus extend the effective viscous theory by Bach and Bruus, J. Acoust. Soc. Am. 144, 766 (2018).\citep{Bach2018} The acoustic temperature field and the thermoviscous boundary layers are incorporated analytically as effective boundary conditions and time-averaged body forces on the thermoacoustic bulk fields. Because it avoids resolving the thin boundary layers, the effective model allows for numerical simulation of both thermoviscous acoustic and time-averaged fields in 3D models of acoustofluidic systems. We show how the acoustic streaming depends strongly on steady and oscillating thermal fields through the temperature dependency of the material parameters, in particular the viscosity and the compressibility, affecting both the boundary conditions and spawning additional body forces in the bulk. We also show how even small steady temperature gradients ($\sim 1\, \SIK/\SImm$) induce gradients in compressibility and density that may result in very high streaming velocities ($\sim 1\,\SImm/\SIs$) for moderate acoustic energy densities ($\sim 100\,\SIJ/\SIm^3$).
\end{abstract}
\maketitle

\section{Introduction}
\seclab{intro}

Modeling and simulation is important for designing microscale acoustofluidic systems. Traditionally, most models have been purely mechanical, but some include thermal effects, such as in the studies of the acoustic radiation force acting on suspended microparticles \citep{Doinikov1997, Danilov2000, Karlsen2015} and of acoustic streaming in rigid cavities. \citep{Rednikov2011, Muller2014}

Here, we focus on acoustic streaming, where recent developments in the field points to the necessity of making a full thermoviscous analysis. Karlsen \textit{et~al.} introduced the acoustic body force acting on a liquid governed by solute-induced gradients in the compressibility and density of the liquid.\citep{Karlsen2016} This force has explained the iso-acoustic focusing of mircoparticles, \citep{Augustsson2016} patterning of concentration profiles, \citep{Karlsen2017} and suppression of acoustic streaming. \citep{Karlsen2018, Qiu2019}
Simultaneously, Bach and Bruus developed the effective theory for pressure acoustics and streaming in elastic cavities,\citep{Bach2018} in which the viscous boundary layer was solved analytically and imposed as an effective boundary condition to the bulk field. This model has enabled simulations of cm-sized three-dimensional (3D) acoustofluidic systems, \citep{Skov2019, Skov2019b} with hitherto prohibitive computational costs, and it has provided a deeper insight in the physics of boundary- and bulk-induced streaming, but without thermal effects.\citep{Bach2019}

In this work, we combine our previous work on thermoviscous streaming in rigid systems, \citep{Muller2014} thermoviscous potential theory, \citep{Karlsen2015} the theory of pressure acoustics with viscous boundary layers and streaming in curved elastic cavities, \citep{Bach2018} and the 3D numerical modeling of acoustofluidic systems using the latter theory, \citep{Skov2019} and develop  an \textit{effective thermoviscous theory} for a fluid-filled cavity embedded in an elastic solid. The theory includes both steady and acoustic temperature fields for pressure acoustics with thermoviscous boundary layers and for streaming with thermoviscous body forces.
In \secref{Model} we set up the basic theory and model assumptions. In Sections~\secnoref{zeroth}-\secnoref{second}, the governing equations and boundary conditions are derived from the theory for the zeroth, first, and second order in the acoustic perturbation, respectively. In \secref{examples}, the theory is implemented in a numerical model, which is then used in two examples to show the nature and importance of thermal effects in acoustofluidics. Finally, we conclude in \secref{conclusion}.

\section{Basic theory and model assumptions}
\seclab{Model}
We consider an acoustofluidic device consisting of an elastic solid containing a microchannel filled with a thermoviscous Newtonian fluid and actuated by an piezoelectric transducer at a single frequency in the $\SIMHz$ range. This time-harmonic actuation establishes an acoustic field in the system, which in the fluid, by the internal dissipation and hydrodynamic nonlinearities, results in a time-averaged response that leads to acoustic streaming.

\subsection{Governing equations}
In this work, unlike prior work,\citep{Skov2019} we leave the piezoelectric transducer out of the analysis, and only represent it by an oscillating displacement condition on part of the surface of the elastic solid. The response of the fluid embedded in the elastic solid to this oscillating-displacement boundary condition is controlled by the hydro-, elasto-, and thermodynamic governing equations of the coupled thermoviscous fluid and elastic solid.

The linear elastic solid is described in the Lagrangian picture by the fields of the density $\rho$, the displacement $\uuu$, and the temperature $T$. There are also associated eight material parameters: the longitudinal and transverse sound speeds $\cL$ and $\cT$, the thermal conductivity $\kth$, the specific heat $\cp$, the ratio of specific heats  $\gamma = c_p/c_v$, the thermal expansion coefficient $\alfP$, and the isentropic and isothermal compressibilities $\kapS$ and $\kapT$. The velocity field is given as the time derivative of the displacement field $\vvv^\mr{sl} =\pp_t \uuu$, so no advection occur, and the governing equations are the transport equations of the momentum density $\rho\pp_t\uuu$ and temperature $T$,\citep{Landau1986, Karlsen2015}
 \bsubal{solid_gov_equ}
 \eqlab{solid_momentum}
 \rho  \ppsqr_t\uuu &= \nablabf \cdot \sigmabf,
 \\
 \eqlab{solid_energy}
 \pp_t T + \frac{(\gamma-1)}{\alfP}\pp_t (\nablabf \cdot \uuu)
 &= \frac{\gamma}{\rho\cp}\nablabf \cdot (\kth\nablabf T),
 \esubal
where $\sigmabf$ is the stress tensor, which for isotropic solids is,
 \bsubal{stress_solid}
 \sigmabf &= -\frac{\alfP}{\kapT} (T-T_0) \III + \taubf,
 \\
 \taubf   &= \rho\cTsqr \Big[\nablabf\uuu + (\nablabf\uuu)^\dagger\Big]
             + \rho\big(\cLsqr-2\cTsqr\big)(\nablabf \cdot \uuu) \III.
 \esubal

The fluid is described in the Eulerian picture by the fields of the density $\rho$, the pressure $p$, the velocity $\vvv$, the temperature $T$, and the energy per mass unit $\epsilon$, and by eight material parameters: the dynamic and bulk viscosity $\eta$ and $\etaB$, the thermal conductivity $\kth$, the specific heat $\cp$, the thermal expansion coefficient $\alfP$, the ratio of specific heats  $\gamma = c_p/c_v$, and the isentropic and isothermal compressibilities $\kapS$ and $\kapT=\gamma \kapS$. The governing equations are the transport equations for the density of mass $\rho$, momentum $\rho\vvv$, and internal energy~$\rho\epsilon$, \citep{Landau1993, Muller2014, Karlsen2015}
 \bsubal{governing_eq}
 \eqlab{govRho}
 \pp_t \rho &= -\nablabf \cdot(\rho \vvv),
 \\
 \eqlab{govV}
 \pp_t(\rho \vvv) &= \nablabf \cdot (\sigmabf -\rho \vvv \vvv),
 \\
 \eqlab{govT}
 \pp_t\Big( \rho \epsilon + \rho \frac{v^2}{2} \Big)
 &= \nablabf\cdot \Big[\kth \nablabf T +\vvv\cdot \sigmabf
 - \rho \vvv \Big(\epsilon + \frac{v^2}{2}  \Big) \Big] + P,
 \esubal
where $P$ is the external heat power density, and $\sigmabf$ is the stress tensor,
 \bsubal{stress}
 \sigmabf &= -p \III + \taubf,
 \\
 \taubf &= \eta \Big[\nablabf \vvv +(\nablabf \vvv)^\dagger\Big]
 +\Big(\etaB - \frac23 \eta\Big)(\nablabf \cdot \vvv) \III.
 \esubal

Pressure and temperature are related to the internal energy density by the first law of thermodynamics, and to the density by the equation of state, \citep{Landau1980, Muller2014, Karlsen2015}
 \bsubal{thermodynamics}
 \eqlab{thermoFirstLaw}
 \rho \dm\epsilon &= (\rho\cp -\alfP p)\: \dm T - (\kapT p+\alfP T)\: \dm p
 \\
 \eqlab{eqofstate}
 \dm \rho &= \rho\kapT\: \dm p - \rho\alfP\: dT
 \esubal
The thermodynamics also shows up in the temperature and density dependency\citep{Muller2014} of any material parameter $q$,
 \beq{water_property}
 \dm q = \Big(\frac{\pp q}{\pp T}\Big)_{\rho} \dm T
 + \Big(\frac{\pp q}{\pp \rho}\Big)_{T} \dm \rho.
 \eeq
The temperature sensitivity of each parameter is quantified by the dimensionless quantity $a_q = \frac{1}{\alfP q} \big(\frac{\pp q}{\pp T}\big)_\rho$,
 \beq{a_water}
 \begin{array}{rcrrcrrcr}
 a_\rho & = &-1,  & \quad
 a_\eta &=& -89, & \quad
 a_{\etaB} &=& -100, \\
 a_{\kth} &=& 11, & \quad
 a_{\alfP} &=& 145, & \quad
 a_{\kapS} &=& -10,
 \end{array}
 \eeq
where the values are for water at $T=25\, \SICel$. \citep{Muller2014} The temperature dependency of the parameters implies that thermal gradients may induce gradients in, say, density and compressibility. This leads to the appearance of the inhomogeneous acoustic body force $\fffac$ introduced in acoustofluidics for solute-induced gradients by Karlsen \textit{et al.} \citep{Karlsen2016}

\subsection{Acoustic actuation and perturbation expansion}
Following Ref.~\onlinecite{Bach2018}, we actuate time-harmonically with angular frequency $\omega$ by a displacement of a surface, so an element at equilibrium position $\sss_0$, at time $t$ will have the position $\sss(\sss_0,t) = \sss_0 + \sss_1(\sss_0)\ee^{-\iot}$. For models containing only a fluid, the displacement will be on the fluid boundary, whereas for models containing both a fluid and a solid domain, the actuation is on the solid boundary. For models including the piezoelectric transducer driving the system, the actuation parameter is the applied voltage.\cite{Skov2019} However, this is not included in this work.

The acoustic response to the actuation parameter $\sss_1$ is linear, and the resulting fields will be complex fields $Q_1(\rrr) \ee^{-\iot}$, the so-called first-order fields with subscript $"1"$. The non-linearity of the governing equation results in higher order responses to the actuation. We are only interested in the time-averaged second-order response and define $Q_2(\rrr) =\avr{Q_2(\rrr,t)} = \frac{\omega}{2\pi} \int_0^{\frac{2\pi}{\omega}} Q_2(\rrr,t)\, \dm t$. A time-average of a product of two first-order fields is also a second-order term, written as $\avr{A_1 B_1}=\frac{1}{2} \re \big[ A_1 B_1^*\big]$, where the asterisk denote complex conjugation. Thus, a given field $Q(\rrr,t)$ in the model, such as density $\rho$, temperature $T$, pressure $p$, velocity $\vvv$, displacement $\uuu$, and stress $\sigmabf$, is written as the sum of the unperturbed field, the acoustic response, and the time-averaged response,
 \beq{Qfield}
 Q(\rrr,t) = Q_0(\rrr) + Q_1(\rrr)\,\ee^{-\iot} + Q_2(\rrr).
 \eeq
Similarly, through their dependency on temperature and density, all material parameters, such as thermal conductivity $\kth$, compressibility $\kappa$, and (for liquids) viscosity $\eta$, are written as exemplified by the viscosity,
 \bsubal{viscosity}
 \eta(\rrr,t) &= \eta_0(T_0) +\eta_1(T_1,\rho_1)\,\ee^{-\iot},
 \\
 \eta_1(T_1, \rho_1)&= \Big(\frac{\pp \eta}{\pp T}\Big)_{T_0} T_1(\rrr) + \Big(\frac{\pp \eta}{\pp \rho}\Big)_{T_0} \rho_1(\rrr).
 \esubal

\subsection{Separation of length scales}
\seclab{model_separation}

Acoustofluidic systems exhibits dynamics on two lengthscales, set by the acoustic wavelength and  the thermoviscous boundary layer width. The boundary conditions on the temperature, heat flux, velocity, and stress at a fluid-solid interface result in the appearance of a thermal boundary layer (in fluids and solids) of width $\delt$ and in a viscous boundary layer (in fluids only) of width $\dels$, localized near fluid-solid interfaces. Their dynamically-defined widths, jointly referred to as $\delta$, are small compared to a typical device size or wavelength $d$, so $\delta \ll d$,\citep{Karlsen2015}
 \beq{boundary_thickness}
 \dels = \sqrt{\frac{2 \nu_0}{\omega}}, \qquad
 \delt = \sqrt{\frac{2 \DthO}{(1-X)\omega}} \approx \sqrt{\frac{2 \DthO}{\omega}},
 \eeq
where $X = 0$ for fluids and $X = (\gamma-1)\frac{4\cTsqr}{3\cLsqr}  \lesssim 0.01$ for solids, $\nu_0=\frac{\eta_0}{\rho_0}$, and $\DthO =\frac{\kthO}{\rho_0 \cpO}$.  Typically, $\delt \lesssim \dels \lesssim 500~\SInm$, which is more than two orders of magnitude smaller than $d \sim 100~\SIum$. In this paper, the various fields are decomposed into a bulk field ($d$) and a boundary-layer field ($\delta$) that are connected by the boundary conditions. In \figref{boundary}, this decomposition is sketched near the fluid-solid boundary for the acoustic temperature field $T_1$. Also shown are the boundary-layer widths $\dels$ and $\delt$  together with the instantaneous position $\sss(t) =\sss_0 +\sss_1(\sss_0,t)$ of the oscillating boundary.

\subsection{Boundary conditions}
In the usual Lagrangian picture,\citep{Bach2018}  an element with  equilibrium position $\sss_0$ in an elastic solid has at time $t$ the position $\sss(\sss_0,t) = \sss_0 + \sss_1(\sss_0)\ee^{-\iot}$ and velocity  $\VVV^0 = -\pp_t\sss = \VVV _1^0(\sss_0)\: \ee^{-\ii \omega t}$ with $\VVV _1^0(\sss_0) = -\ii\omega\sss_1(\sss_0)$. On the solid-fluid interface, the no-slip condition applies, so the velocity of the solid wall at a given time and position must equal the Eulerian-picture fluid velocity $\vvv^\mr{fl}$,
 \beq{noslip_boundary_condition}
 \vvv^\mr{fl}(\sss_0 +\sss_1 \ee^{-\ii \omega t},t) = \VVV _1^0(\sss_0)\:\ee^{-\iot}.
 \eeq
This boundary condition must be obeyed separately for the first- and second-order fields (subscript 1 and 2, respectively), so a Taylor expansion yields\citep{Bach2018}
 \bsubal{noslip_condition}
 \eqlab{no_slip_first}
 \vvv_1(\sss_0) &= \VVV_1^0(\sss_0),
 \\
 \eqlab{no_slip_second}
 \vvv_2(\sss_0) &= -\avr{(\sss_1\cdot \nablabf) \vvv_1}\big|_{\sss_0}
 = -\frac{1}{\omega}\avr{(\ii\VVV_1^0\cdot \nablabf) \vvv_1}\big|_{\sss_0}.
 \esubal
At position $\sss_0$ on the fluid-solid interface with surface normal $\nnn$, also the stress $\sigmabf = \sigmabf_1 + \sigmabf_2$ must be continuous in the first- and second-order contributions $\sigmabf_1$ and $\sigmabf_2$ separately,
 \bsubal{stress_condition}
 \eqlab{stress_first}
 \sigmabf^\mr{sl}_1(\sss_0) \cdot \nnn
 &=
 \sigmabf^\mr{fl}_1(\sss_0)\cdot \nnn,
 \\
 \eqlab{stress_second}
 \sigmabf^\mr{sl}_2(\sss_0)\!\cdot\! \nnn  & =
 \sigmabf^\mr{fl}_2(\sss_0)\!\cdot\! \nnn
 + \avr{(\sss_1\!\cdot\!\nablabf)\sigmabf^\mr{fl}_1(\sss_0)
 \!\cdot\! \nnn}\big|_{\sss_0}.
 \esubal
Here, the thermal effects enter through the temperature dependency of
the viscosity parameters $\eta$ and $\etaB$, see \eqsref{stress}{viscosity}.

\begin{figure}[!t]
 \centering
 \includegraphics[width=\columnwidth]{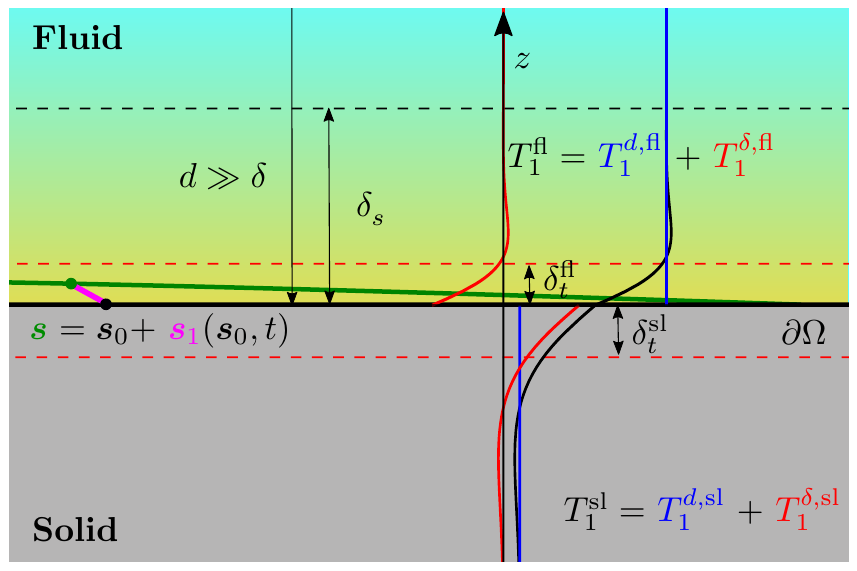}
 \caption[]{\figlab{boundary}
Sketch of the fields at the fluid-solid interface. $\sss_0$ is the equilibrium position of the interface $\pp \Omega$,  $\sss_1$ the time-dependent displacement away from $\pp\Omega$, and $\sss = \sss_0+\sss_1$ the instantaneous position. The dashed lines represent the viscous and thermal boundary-layer widths  width $\dels$ (black) and $\delt$ (red) in the solid and fluid. $\delta$ without a subscript refers to either $\dels$ or $\delt$, and $d$ refers to the bulk lengthscale, so $\delt\lesssim\dels \sim \delta \ll d$. The temperature $T_1^\mr{xl}$ (black) is the sum of a bulk field $T_1^{d,\mr{xl}}$ (blue) and a boundary-layer field $T_1^{\delta\mr{xl}}$ (red).
}
 \end{figure}

Similarly, the temperature $T = T_0+ T_1 + T_2$ must be continuous across the solid-fluid interface in each order separately,
 \bsubal{temp_condition}
 \eqlab{temp_first}
 T_i^\mr{sl}(\sss_0) &= T_i^\mr{fl}(\sss_0), \quad i = 0,1,
 \\
 T_2^\mr{sl}(\sss_0) &=
 T_2^\mr{fl}(\sss_0) + \avr{\sss_1\cdot \nablabf T^\mr{fl}_1}\big|_{\sss_0}.
 \esubal
Also the heat flux $\nnn\cdot(-\kth \nablabf T)$ must be continuous across the interface,
 \beq{heat_flux_boundary_condition}
 \kthsl \nnn \cdot \nablabf T^\mr{sl}(\sss_0,t) =
 \kthfl \nnn \cdot \nablabf T^\mr{fl}(\sss_0 +\sss_1 \ee^{-\ii \omega t},t),
 \eeq
which order by order becomes,
 \bsubal{heatflux_condition}
 \eqlab{heatflux_first}
 &\kthOsl \nnn \cdot \nablabf T_i^\mr{sl}(\sss_0)
 = \kthOfl \nnn \cdot \nablabf T_i^\mr{fl}(\sss_0), \quad i = 0,1,
 \\[2mm]
 \eqlab{heatflux_second}
 &\kthOsl \nnn \cdot \nablabf T_2^\mr{sl} + \kthIIsl \nnn \cdot \nablabf T_0^\mr{sl}
 +\avr{\kthIsl \nnn \cdot \nablabf T_1^\mr{sl}}
 \\
 \nn
 &= \kthOfl \nnn \cdot \nablabf T_2^\mr{fl}(\sss_0)
 + \avr{\kthIfl \nnn \cdot \nablabf T_1^\mr{fl}(\sss_0)}
 \\
 \nn
 &\quad + \kthIIfl \nnn \cdot \nablabf T_0^\mr{fl}(\sss_0)
 +\avr{\sss_1 \cdot \nablabf \big[ \kthOfl \nablabf  T_1^\mr{fl}(\sss_0) \big]\cdot \nnn}
 \\
 \nn
 &\quad+ \avr{\sss_1 \cdot \nablabf \big[ \kthIfl \nablabf  T_0^\mr{fl}(\sss_0) \big]\cdot \nnn}.
 \esubal

\subsection{Range of validity of the model}
\seclab{validity_range}

We briefly discuss the range of validity imposed by the main assumptions. Firstly, perturbation theory is valid when lower-order terms are much larger than and unaffected by higher-order terms, say, $\rhoO \gg \abs{\rhoI}$ and $\abs{\vvv_1}\gg \abs{\vvv_2}$, and when the latter can be neglected in the governing equations. For example, the zeroth-order heat equation~\eqnoref{energy0th} is only valid, if the time scale for advective heat transport $t_\mr{adv}=d_t/\abs{\vvv_2}$ is much longer than that of diffusion $t_\mr{dif}=d_t^2/\DthO$ in a system with characteristic length $d_t$. For $d_t =1\,\SImm$ this requires $\abs{\vvv_2}\ll \frac{\DthO}{d_t}\approx 150\,\SImum/\SIs$.

Secondly, due to low oscillatory advection, we assume  $\nablabf \cdot (q_0 \vvv_1) \approx q_0 \nablabf \cdot \vvv_1$, where $q_0$ is a parameter of the fluid. This requires $\vert q_0 \nablabf \cdot \vvv_1 \vert \gg \vert\nablabf q_0 \cdot \vvv_1\vert$. By using the parameter $a_q$ of \eqref{a_water}, the validity of our theory is limited by,
 \beq{T0_limit}
 \abs{\nablabf T_0 } \ll \bigg|\frac{\kc}{a_\eta \alfPO}\bigg| \approx 5000~\frac{\SIK}{\SImm}.
 \eeq
Here, $a_\eta$ is used as the viscosity has the strongest temperature dependency. In conventional acoustofluidic systems $\abs{\nablabf T_0 } \lesssim 50~\SIK/\SImm \ll 5000~\SIK/\SImm$.

Thirdly, the effective boundary-layer theory requires the boundary-layer width to be much smaller than the bulk wavelength, $k_0\delta \ll 1$, see \secref{model_separation}, which is true for MHz acoustics in water.

\section{Zeroth order: steady background fields}
\seclab{zeroth}
Before turning on the acoustics, $p_0$ is constant, and $\vvv_0 = \zerovec$ in the acoustofluidic system. The temperature $T_0$ is determined by boundary conditions set by the surroundings and the heat power density $P_0$ from given sources and sinks. $T_0$ is governed by the energy conservation~\eqnoref{govT} to zeroth order in the acoustic actuation,
\beq{energy0th}
 0= \nablabf \cdot \Big[ \kthO \nablabf T_0\Big] + P_0.
\eeq
$T_0$ determines the zeroth-order water parameters, such as $\rho_0(T_0)$ and $\eta_0(T_0)$, and thereby affects the resonance frequency and the Q-factor of the acoustofluidic system.

\section{First order: acoustics}
\seclab{first}
For the first-order fields, we solve the viscous and thermal boundary layers analytically, and use these solutions to derive a set of effective boundary conditions for the bulk fields. The analysis is based on our previous work: the governing equations derived in Refs.~\onlinecite{Muller2014, Karlsen2015}, the potential theory derived in Ref.~\onlinecite{Karlsen2015}, and the effective boundary method derived for viscous, but not for thermal, boundary layers in Ref.~\onlinecite{Bach2018}. The result is a model, where we solve for the displacement field $\uuu_1$ in the solid, and for the pressure $p_1$ in the fluid, and both these bulk fields are subject to the effective boundary conditions that implicitly contain the boundary layers. The temperature $T_1$ is incorporated through $p_1$, $\uuu_1$ in the first-order equations and in the effective boundary conditions.

\subsection{Acoustic equations and potential theory for fluids}
\seclab{FirstOrderFluids}
The governing equations for the complex-valued acoustic field amplitudes in a fluid are given in Eq.~(11) of Ref.~\onlinecite{Muller2014}: the mass continuity equation, the momentum equation, and the heat equation, which couple together the pressure $p_1$, the velocity $\vvvI$, and the temperature $T_1$,
 \bsubal{govEquI}
 \eqlab{MconsI}
 &-\ii\omega \alfPO\:\TI + \ii \omega \kapTO\:\pI = \nablabf\cdot\vvvI,
 \\
 \eqlab{PconsI}
 &-\ii\omega\rhoO\: \vvvI = -\nablabf\pI +\beta\etaO\nablabf(\nablabf\cdot\vvvI)+\etaO\Lapl\vvvI,
 \\
 \eqlab{EconsI}
 &-\ii\omega \TI  + \ii\omega (\gamma-1)\frac{\kapsO}{\alfPO}\: \pI = \DthO\Lapl\TI,
 \esubal
where $\beta = \frac{\etaBO}{\etaO} -\frac23$. Following  Ref.~\onlinecite{Karlsen2015}, these equations are solved using potential theory based on the standard Helmholtz decomposition of the velocity field, $\vvvI = \nablabf(\phi_c + \phi_t) + \nablabf\times\vPsi = \vvvId + \vvvIdel$, where $\phi_c$ is the compressional potential, $\phi_t$ is the thermal potential, and $\vPsi$ is the shear vector potential. At the fluid-solid interface $\big|\TIdel\big| \approx \big|\TId \big|$, and combining this with $T_1 = \TId + \TIdel = \frac{\ii(\gamma-1)\omega}{\alfPO c_0^2}\phi_c + \frac{1}{\alfPO\DthO}\phi_t$  with the typical acoustofluidic parameter values inserted, we can deduce $\big|\phi_t\big| \approx (\gamma-1)\frac{\omega\DthO}{c_0^2}\big|\phi_c\big| \approx 10^{-8}\big|\phi_c\big| \ll \big|\phi_c\big|$. From this follows that $p_1 \approx \ii\omega\rhoO(1+\ii\Gamv)\phi_c$, and we replace $\phi_c$, $\phi_t$, and $\vPsi$ by $p_1$, $\TIdel$, and $\vvvIdel$,
 \beq{phicphitPsi}
 p_1 \approx \ii\omega\rhoO(1+\ii\Gamv)\phi_c,
 \;\;
 \TIdel = \frac{\phi_t}{\alfPO\DthO},
 \;\;
 \vvvIdel = \rot\vPsi.
 \eeq
Finally, using the smallness of the damping coefficients, $\Gamv = \frac12(1+\beta)(k_0\dels)^2 \ll 1$ and  $\Gamt = \frac12(k_0\delt)^2 \ll 1$, with $k_0 = \frac{\omega}{c}$, approximate solutions to \eqref{govEquI} are obtained from the potentials solving three Helmholtz equations,
 \bsubalat{Helmholtz}{2}
 \eqlab{Helmholtz_p1}
 \nabla^2 p_1 &= -\kcsqr p_1,
 &\quad \kc &= \frac{\omega}{c_0}(1+\ii\GamOcfl),
 \\
 \eqlab{Helmholtz_T1del}
 \nabla^2 \TIdel  &= -\ktsqr \TIdel,
 & \kt &= \frac{1+\ii}{\delt}(1+\ii\GamOtfl),
 \\
 \eqlab{Helmholtz_v1del}
 \nabla^2\vvvIdel &= -\kssqr\vvvIdel,
 & \ks &= \frac{1+\ii}{\dels}.
 \esubalat
Here, $\GamOcfl = \frac{1}{2}\big[\Gamv + (\gamma-1) \Gamt \big]$ and $\GamOtfl = \frac{\gamma-1}{2}\big[\Gamv - \Gamt\big]$ are the resulting damping coefficients, whereas the complex-valued wave numbers $\ks$ and $\kt$ reveals the existence of the viscous and thermal boundary layers of thickness $\dels$ and $\delt$, respectively, see \figref{boundary}. The full velocity $\vvvI$ and temperature $T_1$ are given by $p_1$, $\vvvIdel$, and $\TIdel$ as,
 \bsubalat{v1T1}{2}
 \eqlab{v1full}
 \vvvI &= \vvvId + \vvvIdel
  \qquad = & \vvv_1^{d,p}  &\; + \vvv_1^{d,T} + \vvvIdel,
 \\
 \eqlab{v1dpdT}
 \vvv_1^{d,p} &= \grad\big[-\ii\mbox{$\frac{1-\ii\Gamv}{\omega\rhoO}$}\:p_1\big],
 \;\;&
 \vvv_1^{d,T} &= \grad\big[\alfPO \DthO T_1^\delta\big],
 \\
 \eqlab{T1flfull}
 T_1 &= \TId + \TIdel, \quad &
 \TId &= (\gamma-1)\frac{\kapsO}{\alfPO}\:p_1.
 \esubalat
Note that both $\vvv_1^{d,p}$ and $\vvv_1^{d,T}$ are gradient fields in the Helm\-holtz decomposition, but  that $\vvv_1^{d,T}$ despite its superscript \qmarks{d} is a boundary-layer field.  Because $T_1$ is split into a bulk and a boundary layer field, the material parameters $q = q_0 + q_1$ are split similarly. For example, the first-order viscosity $\eta_1$ introduced in \eqref{viscosity} (and similar for other material parameters) becomes
 \beq{eta1full}
  \eta_1 =\pp_T\eta_0 \;(\TId + \TIdel) +\pp_\rho \eta_0 (\rhoI^d + \rhoI^\delta)= \eta_1^d + \eta_1^\delta.
 \eeq

\subsection{Acoustic equations and potential theory for solids}
For a linear elastic isotropic solid with density $\rhoO$, longitudinal sound speed $\cL$, and transverse sound speed $\cT$, the governing equations is the linearized form of the momentum and heat equation~\eqnoref{solid_gov_equ} for the displacement field $\uuuI$ and the temperature $T_1$,\citep{Karlsen2015}
 \bsubal{govEquIsl}
 \eqlab{PconsIsl}
 -\omega^2\rhoO\uuuI = & -\frac{\alfPO}{\kapTO}\nablabf\TI
 \nn \\
 &  +
 (\cLsqr-\cTsqr)\nablabf(\nablabf\cdot\uuuI)+\cTsqr\Lapl\uuuI,
 \\
 \eqlab{EconsIsl}
 -\ii\omega \TI  - &\; \ii\omega \frac{\gamma-1}{\alfPO}\: \nablabf\cdot\uuuI
 = \DthO\Lapl\TI.
 \esubal
In analogy with the fluid, the governing equations for the solid are solved by potential theory, again following Ref.~\onlinecite{Karlsen2015}. The displacement field is Helmholtz decomposed as $-\ii\omega\uuuI = \nablabf(\phi_c+\phi_t) + \nablabf\times\vPsi = -\ii\omega(\uuuIL + \uuuIT)$, where $\phi_c$ is the compressional potential, $\phi_t$ is the thermal potential, and $\vPsi$ is the shear vector potential, and where we have used $\vvvIsl = -\ii\omega\uuuI$. Using the same approximations as for the fluid, we have $T_1 = \TId + \TIdel = \frac{\ii(\gamma-1)\omega}{\alfPO c_0^2}\phi_c + \frac{1}{\chi\alfPO\DthO}\phi_t$.  We keep $\phi_c$, but use $\TIdel = \frac{1}{\chi\alfPO\DthO}\phi_t$ instead of $\phi_t$, and $\uuuIT = \rot\vPsi$ instead of $\vPsi$. The solution to \eqref{govEquIsl} is obtained from the potentials solving the following three Helmholtz equations,
 \bsubalat{Helmholtzsl}{2}
 \eqlab{Helmholtzsl_phic}
 \nabla^2 \phi_c &= -\kcsqr \phi_c,
 &\quad \kc &= \frac{\omega}{c_0}(1+\ii\GamOcsl),
 \\
 \eqlab{Helmholtzsl_T1del}
 \nabla^2 \TIdel  &= -\ktsqr \TIdel,
 & \kt &= \frac{1+\ii}{\delt}(1+\ii\GamOtsl),
 \\
 \eqlab{Helmholtzsl_u1tr}
 \nabla^2\uuuIT &= -\kssqr\uuuIT,
 & \ks &= \frac{\omega}{\cT}.
 \esubalat
Here, $c_0^2 = \cLsqr + \frac{\gamma-1}{\rhoO\kapTO}$, $\GamOcsl = \frac{\gamma-1}{2}\chi\Gamt$, and $\GamOtsl = \frac{\gamma^2\Gamt}{8(1-X)}$ are damping coefficients,  $\delt$ and $\Gamma_t$ are given by \eqref{boundary_thickness}, $\chi = 1-\frac{4\cTsqr}{3c^2} \approx \frac12$, and $X = (\gamma-1)\frac{4\cTsqr}{3c^2} \approx \frac{\gamma-1}{2}$. For a solid, only $\TIdel$ is a dampened field confined to the boundary layer, whereas $\phi_c$ and $\uuuIT$ are bulk fields. The transverse waves in fluids and solids are qualitatively different: $\vvvIdel$ cannot propagate in a fluid and is restricted to the boundary layer, whereas $\uuuIT$ can propagate in a solid and is not associated with a boundary layer.
The full displacement $\uuuI$ and temperature $T_1$ are given by $\phi_c$, $\uuuIT$, and $\TIdel$ as,
 \bsubalat{u1T1}{2}
 \eqlab{u1full}
 \uuuI &= \uuuIL + \uuuIT, \qquad&
 \uuuIL &= \frac{\ii}{\omega}\:\nablabf \phi_c,
 \\
 \eqlab{T1slfull}
 T_1 &= \TId + \TIdel, \qquad &
 \TId &=  \frac{\ii(\gamma-1)\omega}{\alfPO c_0^2}\phi_c .
 \esubalat

The explicit expression for the stress tensor $\sigmabf^\mr{xl}_1$ in the fluid (xl = fl) and in the solid (xl = sl) can be formulated jointly in potential theory as\citep{Karlsen2015}
\bal
 \eqlab{sigma1xl}
 \sigmabf_1^\mr{xl}
 &= -p_1^\mr{xl} \mathbf{I} +\eta_0^\mr{xl}
 \big[(2\kcsqr-\kssqr)\phic + (2\ktsqr-\kssqr)\phit\big] \mathbf{I}
 \nn
 \\
 &\quad + \eta_0^\mr{xl} \big[\grad \vvvI^\mr{sl} + (\grad \vvvI^\mr{sl})^\dagger\big],
 \eal
where in the solid $p_1^\mr{sl} = 0$, $\eta_0^\mr{sl} = \frac{\ii}{\omega}\:\rho_0 \cT^2$, $\vvvI^\mr{sl} = -\ii\omega\uuuI$.

\subsection{The thermal boundary layer}
The temperature fields $T_1^{\delta,\mr{xl}}$ in the fluid (xl = fl) and the solid (xl = sl) are given by \eqsref{Helmholtz_T1del}{Helmholtzsl_T1del}. Following Ref.~\onlinecite{Bach2018} with $x$ and $y$ parallel to the interface and $z$ perpendicular, an analytical solution can be found using the thin-boundary-layer approximation $\Lapl \approx \pp_z^2$ in these equations in combination with the condition that the field decays away from the boundary,
 \bsubalat{T1_boundary}{2}
 \TIdelfl(x,y,z)
 &= \TIdelOfl(x,y)\: \ee^{\ii \kt^\mr{fl} z},\; &&\text{ for } z > 0,
 \\
 \TIdelsl(x,y,z)
 &= \TIdelOsl(x,y)\: \ee^{-\ii \kt^\mr{sl} z},\; &&\text{ for } z < 0.
 \esubalat
The amplitude of the boundary fields $\TIdelOfl(x,y)$ and $\TIdelOsl(x,y)$ are determined by the boundary conditions \eqsnoref{temp_first}{heatflux_first} as follows. The normal vector $\nnn = -\een_z$ points away from the fluid, so $\nnn \cdot \nablabf = - \pp_z$, and we obtain
 \bsubal{thermal_boundary_conditions}
 \eqlab{thermal_boundary_conditions_T}
 \TIdelOfl& = \TIdelOsl - \Delta T_1^{d0},
 \\
 \eqlab{thermal_boundary_conditions_nablaT}
 \kthOfl \pp_z  \TIdelfl
 &= \kthOsl \pp_z  \TIdelsl,\;\text{ for } z = 0,
 \esubal
where $\Delta T_1^{d0} = -\big(\TIdOfl - \TIdOsl\big)$.
From \eqref{thermal_boundary_conditions_nablaT} follows the relation,
 \beq{Tdeltas}
 \TIdelOfl =
 - \frac{\kthOsl\kt^\mr{sl}}{\kthOfl\kt^\mr{fl}}\: \TIdelOsl
 = - \tilde{Z}\: \TIdelOsl,
\eeq
where $\tilde{Z} =  Z^\mr{sl}/Z^\mr{fl}$ is the ratio of $Z=\kthO \kt=\sqrt{\kthO \cpO \rhoO}$ of the solid and the fluid, respectively. Combining \eqsref{thermal_boundary_conditions_T}{Tdeltas} leads to the final expression for the boundary-layer fields,
 \bsubal{T1_delta}
 \eqlab{T1_delta_fl}
\TIdelfl(x,y,z)
 &= -\frac{\tilde{Z}}{1+ \tilde{Z}}\: \Delta T_1^{d0}(x,y)\: \ee^{\ii \kt^\mr{fl} z},
 \\
 \eqlab{T1_delta_sl}
 \TIdelsl(x,y,z)
 &= +\frac{1}{1+ \tilde{Z}}\: \Delta T_1^{d0}(x,y)\: \ee^{-\ii \kt^\mr{sl} z}.
\esubal

\subsection{The viscous boundary layer}
The viscous boundary layer exists only in the fluid, since in the solid both $\uuuIL$ and $\uuuIT$ are bulk fields. The velocity field in the fluid is given in \eqref{v1full} as $\vvvI = \vvvId + \vvvIdel$, where $\vvvId$ depends on the bulk field $p_1$ and the boundary field $\TIdel$. The boundary field $\vvvIdel$ is given by the Helmholtz equation~\eqnoref{Helmholtz_v1del}, to which an analytical solution can be found using the thin-boundary-layer approximation $\Lapl \approx \pp_z^2$ in combination with the condition that the field decays away from the boundary,\citep{Bach2018}
 \beq{v1delta}
 \vvvIdel = \vvvIdelO(x,y)\: \ee^{\ii k_s z}.
 \eeq
The amplitude $\vvvIdelO$ of the boundary field is determined by the no-slip condition~\eqnoref{no_slip_first},
\beq{vshear_boundary}
\vvvIdelO = \VVV_1^0 - \vvvIdO =  -\ii\omega\uuuIO  - \vvvIdO.
\eeq

\subsection{The effective boundary condition for the velocity}
Given the analytical solutions of the three boundary-layer fields, we only need to numerically solve the three bulk fields, namely $\phi_c$ and $\vPsi$ in the solid and $\phi_c$ in the fluid, or equivalently, the displacement $\uuu_1$ in the solid and the pressure $p_1$ in the fluid. Therefore, we set two effective boundary conditions on these bulk fields using the analytical solutions for the boundary-layer fields: One effective boundary condition on the displacement $\uuu_1$ in the solid derived from the condition on the stress, and another on the pressure in the fluid.

First, from the no-slip condition~\eqnoref{no_slip_first}, we derive the boundary condition for the first-order pressure field $p_1$, which takes the viscous and thermal boundary-layer effects into account through terms with $\ks$, $\kt$, and $T_1^{\delta 0}$. We express  the compressional velocity $\vIdOflz$ on the fluid-solid interface through the no-slip condition~\eqnoref{vshear_boundary}, then use the incompressibility condition on the boundary-layer velocity, $\ii\ks\vIdelOflz + \divop\vvvIdelOfl = 0$, to get rid of the $z$-component $\vIdelOflz$, and finally introduce the bulk fields in the fluid,
 \bal
 \vIdOflz
 &= \vIdOslz  - \vIdelOflz = \vIdOslz - \frac{\ii}{\ks}\divop\vvvIdelOfl
 \nn
 \\
 \eqlab{noslip_boundary}
 &= \vIdOslz-\frac{\ii}{\ks}\divop \Big[\vvvIdOsl - \vvvIdOfl\Big]
 \\
 \nn
 &= \Big[\vIdOslz\!-\!\frac{\ii}{\ks} \nablabf\scap\vvvIdOsl \Big] +
 \frac{\ii}{\ks} \Big[\nablabf\scap\vvvIdfl\!-\! \pp_z \vIdflz\Big]_{z=0}.
 \eal
Combining \eqsref{MconsI}{T1flfull}, we obtain
 \bsuba{rewrite_v1_terms}
 \beq{div_v1d}
 \divop \vvv_1^\mathrm{d} =
 \ii\frac{(1-\ii \Gams)\kc^2}{\omega \rho_0}  p_1 - \ii \omega \alfPO T_1^\delta.
 \eeq
Then using \eqref{v1full}, we write $ \vIdOflz$ and $\pp_z \vIdflz$ evaluated at the solid-fluid interface at $z=0$, and arrive at
 \bal
 \vIdOflz
 & =  -\frac{\ii}{\omega \rho_0} (1-\ii\Gams) \pp_z p_1
 + \alfPO \DthO  \pp_z T_1^\delta,
 \\
 \pp_z  \vIdflz
 & =  -\frac{\ii}{\omega \rho_0} (1-\ii\Gams) \pp_z^2 p_1
 + \alfPO \DthO \pp_z^2 T_1^\delta.
 \eal
 \esuba
Inserting \eqsref{rewrite_v1_terms}{no_slip_first} into \eqref{noslip_boundary} leads to the final form of the effective boundary condition on $p_1$,
 \bsub
 \eqlab{p1u1BCfinal}
 \bal
 \pp_z p_1
 &= \ii\frac{\omega \rho_0}{1-\ii \Gams} \big(V_{1z}^0- \frac{\ii}{\ks} \divop \VVV_1^0 \big)
 -\frac{\ii}{\ks} \big(\kc^2 +\pp_z^2\big) p_1
 \nn
 \\
 \eqlab{p1BCeffective}
 & \quad + \frac{\ii}{\kt} \frac{\alfPO}{\kapTO} \kOsqr\, T_1^{\delta 0},\; \text{ for } z = 0.
 \eal
The first two terms on the right-hand were derived by Bach and Bruus, \citep{Bach2018} whereas the last term is a new correction due to the thermal boundary layer. For $T_1^d \approx T_1^\delta$ at $z=0$, this thermal correction is of the order $\frac{\gamma-1}{\kt}\:\kc^2 \:p_1$. We emphasize, that although formulated as an effective boundary condition on the pressure gradient,  \eqref{p1BCeffective} is the no-slip velocity condition.

\subsection{The effective boundary condition for the stress}
Next, using the explicit expressions for $\sigmabf_1^\mr{sl}$ and  $\sigmabf_1^\mr{fl}$, we turn to the stress boundary condition~\eqnoref{stress_first}, the continuity of the stress $\sigmabf_1$ across the fluid-solid interface, $\sigmabf_1^\mr{sl}\cdot \eee_z = \sigmabf_1^\mr{fl}\cdot \eee_z$. For the fluid, we use that $\ks \gg \kc$, $\abs{\phic} \gg \abs{\phit}$, and $\abs{\pp_z \vvv_1^\delta} \gg \abs{\nabla \vvv_1^d}$ in \eqref{sigma1xl}, and find
 \beq{stress_fluid}
 \sigmabf_1^\mr{fl}\cdot \eee_z =
 -p_1\eee_z +\ii\ks\eta_0 \Big[\vvvIdOsl + \frac{\ii}{\omega \rho_0}\nablabf p_1 \Big]_{\sss_0}.
\eeq
For the solid, we neglect in \eqref{sigma1xl} the derivative $\pp_\parallel \phit$ along the surface, as it is a factor $\Gamt$ smaller than $\pp_\parallel \phic$. The remaining $\phit$-dependent boundary-layer terms cancel out, leaving only the bulk-term part $\sigmabf_1^{d,\mr{sl}}$ of $\sigmabf_1^\mr{sl}$. The resulting effective stress boundary condition becomes,
 \beq{stress_eff_boundary}
 \sigmabf_1^{d,\mr{sl}}\cdot \eee_z = \sigmabf_1^\mr{fl}\cdot \eee_z.
 \eeq
 \esub
As the thermal boundary-layer fields do not enter, this expression is identical to the effective boundary condition for the stress derived in Ref.~\onlinecite{Bach2018}.

\section{Second order: Acoustic streaming}
\seclab{second}
For the second-order fields in the fluid, we follow \eqref{Qfield} and consider only the time averaged fields, namely the velocity $\vvv_2$, pressure $p_2$ and stress $\sigmabf_2$. The temperature field $T_2$ does not enter the second-order continuity or Navier--Stokes equation, so we drop the heat equation. The first-order temperature field $T_1$ enters the equations through the  material parameters of the fluid,
 \bsubal{2nd_order_eq}
 \eqlab{2nd_order_ContEqu}
 0 &= -\nablabf \cdot (\rho_0 \vvv_2)+\rhoac
 \\
 0 &= -\nablabf p_2+\nablabf\cdot\taubf_2  + \fachat \\
 \taubf_2 &= \eta_0 \big[ \nablabf \vvv_2 + (\nablabf \vvv_2)^\dagger\big] + \big[\etaBO-\tfrac23 \eta_0\big] (\nablabf \cdot \vvv_2)\,\III,
 \\
 \eqlab{v20def}
 \vvv_2^{0} &=
 -\frac{1}{\omega}\avr{\big(\ii\VVV_1^0 \cdot \nablabf\big) \vvv_1}\Big|_{\rrr = \sss_0}.
 \esubal
Here, the excess density rate $\rhoac$ and the acoustic body force $\fachat$ are time-averaged products of fast varying first-order fields, which, assuming that $\rho_0 \vvv_1\gg \rho_1 \vvv_0$ as is true for typical acoustofluidic devices, are given by
 \bsubal{acoustic_contribution}
 \eqlab{excess_density}
 \rhoac  &=- \nablabf\cdot \avr{\rho_1 \vvv_1},
 \\
 \eqlab{acoustic_body_force}
 \fachat &= \nablabf \cdot\big[-\rho_0\avr{\vvv_1 \vvv_1}+ \taubf_{11}\big],
 \\
 \taubf_{11} &= \avr{\eta_1\big[\nablabf\vvv_1 \!+\! (\nablabf\vvv_1)^\dagger\big]
 +\big[\eta_1^\mr{b} \!-\!\tfrac23\eta_1\big](\divop\vvv_1)\III}.
 \esubal
The slowly varying second-order fields are split up in a bulk field (superscript \qmarks{$d$}) and a boundary field (superscript \qmarks{$\delta$}) according to their  response to the boundary and bulk part of the acoustic force $\fachat = \fachat^d +\fachat^\delta$, and they are coupled by the boundary conditions,
 \bsubalat{2nd_order_fields}{2}
 p_2 &= p_2^d + p_2^\delta,
 \qquad &
 \vvv_2 &= \vvv_2^d+\vvv_2^\delta,
 \\
 \taubf_2&= \taubf_2^d +\taubf_2^\delta,
 \qquad &
 \taubf_{11} &= \taubf_{11}^d+\taubf_{11}^\delta.
 \esubalat
Note that in contrast to the first-order fields, this is not a Helmholtz decomposition: by definition, a second-order boundary-layer field \qmarks{$\delta$} contains at least one first-order boundary-layer field.
The computation strategy for second-order streaming is similar to the one for first-order acoustics: (1) find analytical solution to the boundary layers, (2) formulate effective boundary conditions, and (3) solve the bulk fields with the effective boundary conditions. This decomposition enables simulations of the bulk fields without resolving the boundary-layer fields.

\subsection{Short-range boundary-layer streaming}
The short-range part \qmarks{$\delta$} of \eqref{2nd_order_eq} is given by the short-range part of the second-order fields as well as all source terms containing at least one boundary-layer field,
\bsubal{2nd_delta_eq}
 0 &= \nablabf \cdot \big( \rho_0 \vvv_2^\delta\big) + \rhoac^\delta,
 \\
 \eqlab{v2delGov}
 0& = -\nablabf p_2^\delta +\nablabf\cdot \taubf_2^\delta + \fachat^\delta,
 \\
 &\qquad \mr{where }\; \vvv^\delta_2 \rightarrow 0 \;\mr{ as }\; z\rightarrow \infty.
\esubal
At the boundary, the advection term can be neglected compared to the viscous term, because of the large gradients induced by the small lengthscale $\delta$. The thermal boundary layer $T_1^\delta$ and the associated boundary-layer velocity $\vvv_1^{d,T}$ introduce a correction $\vvv_{2}^{\delta,T}$ to the purely viscous boundary-layer term $\vvv_{2}^{\delta,p}$ computed in Ref.~\onlinecite{Bach2018},
 \beq{v2dpdT}
 \vvv_{2}^\delta =  \vvv_{2}^{\delta,p} +  \vvv_{2}^{\delta,T}.
 \eeq
In the parallel component of $\vvv_2^\delta$, the pressure field can be neglected,  because $\pp_\parallel p_2^\delta \ll \eta_0\pp_z^2 \vvv_{2\parallel}^\delta$.\citep{Bach2018}
Thus combining \eqsref{acoustic_body_force}{v2delGov}, the parallel component of the short-range velocity field $\vvv_{2}^{\delta,T}$ obeys,
 \bal\eqlab{JSB39a}
 \nu_0 \pp_z^2 \vvv_{2,\parallel}^{\delta,T} &= \Big[
 \nablabf \cdot \big\langle\vvv_1^\delta \vvv_1^{d,T}+\vvv_1^{d,T} \vvv_1^\delta+\vvv_1^{d,p} \vvv_1^{d,T}\nn\\&+ \vvv_1^{d,T} \vvv_1^{d,p} +\vvv_1^{d,T} \vvv_1^{d,T}\big\rangle
 -\frac{1}{\rho_0}\nablabf \cdot \taubf_{11}^{\delta}\Big]_\parallel.
 \eal
Here, $\taubf_{11}^{\delta}$ depends on $T_1$ through $\eta_1(T_1)$, whereas the velocity $\vvv_1^{d,T}$, given in \eqnoref{v1full}, depends on the thermal boundary layer $T_1^\delta$. From \secref{first} and in particular Eqs.\ \eqnoref{v1T1}, \eqnoref{T1_boundary}, and \eqnoref{v1delta} follow the relations $\nablabf\scap\vvv_1^\delta =0$, $\big|\vvv_{1,\parallel}^\delta\big| \approx \big|\vvv_{1,\parallel}^{d}\big|$, $\big|v_{1,z}^\delta\!\big| \approx (k_c\dels)\big|v_{1,z}^{d}\big|$,
$\big|T_1^{\delta}\big| \approx \big|T_1^d\big|$,
$\divop\vvv_1^{d,T} \approx (\gamma\!-\!1) \divop\vvv_1^{d,p}$,
$\abs{\vvv_{1,z}^{d,T}}\approx (\gamma-1) (\kc\delt) \abs{\vvv_{1,z}^{d,p}}$,
$\abs{\vvv_{1,\parallel}^{d,T}}\approx (\gamma\!-\!1) (\kc\delt)^2 \abs{\vvv_1^{d,p}}$, and $\vvv_1^{d,T} = \alfPO \DthO\,\nablabf T_1^\delta$. To lowest order in $\kc\delta \ll 1$ (involving  $\pp_z T_1^\delta$ and $\pp_z\vvv_1^\delta$, respectively), these relations combined with time averaging $\avr{a_1b_1} = \frac12\re\!\big[a_1b_1^*\big]$ change \eqref{JSB39a} to
 \bal\eqlab{JSB39a_leading}
 \nu_0 \pp_z^2 \vvv_{2,\parallel}^{\delta,T} &=
 \Big[\avr{\big(\partial_z \vvv_1^\delta\big) v_{1,z}^{d,T}}
 + \avr{ \big(\vvv_1^\delta + \vvv_1^{d,p}\big) \big(\partial_zv_{1,z}^{d,T}\big)}
 \nn
 \\
 & \quad
 - \frac{1}{\rho_0}\Big(
 \avr{(\pp_z\eta_1^\delta)\pp_z\vvv_1^\delta} + \avr{(\eta_1^\delta + \eta_1^d)\Lapl\vvv_1^\delta}
 \Big)\Big]_\parallel
 \nn
 \\
 & = \frac12\re\!\bigg[\frac{2 \alfPO \DthO}{\delt^2}
 \bigg(\frac{\delt+\ii\dels}{\dels} \vvv_1^\delta
 + \ii \vvv_1^{d,p}\bigg)\, T_1^{\delta*}
 \nn
 \\
 &\qquad \quad -\frac{2}{\rho_0\dels^2}
 \bigg(\frac{\dels+\ii\delt}{\delt}\eta_1^\delta + \ii\eta_1^d\bigg)\,\vvv_1^{\delta *}
 \bigg]_\parallel.
 \eal
The integration of \eqref{JSB39a_leading} after $z$ twice, is facilitated by using the analytical forms \eqnoref{T1_boundary} and  \eqnoref{v1delta} for $\vvv_1^{d,T}$, $T_1^\delta$, and $\vvv_1^\delta$, and by noting that in the boundary layer $\eta_1^d \approx \eta_1^{d0}\!+\!z\pp_z\eta_1^d \approx (1\!+\!\kc\dels)\eta_1^{d0} \approx \eta_1^{d0}$ and similarly $\vvv_1^d \approx \vvv_1^{d0}$,
 \bsubalat{Q1delta}{2}
 \vvv_1^\delta &= \vvv_1^{\delta0}(x,y)\, q(z),\,
 &\text{ with }&\; q(z) = \ee^{\ii\ks z},
 \\
 T_1^\delta &= T_1^{\delta 0}(x,y)\, r(z),
 &\text{ with }&\; r(z) = \ee^{\ii\kt z},
 \\
 \eta_1^\delta &= \eta_1^{\delta0}(x,y)\, r(z),\;
 &
 \\
 \eta_1^d &\approx \eta_1^{d0}\quad \text{ and }
 &
 \vvv_1^{d,p}  \approx &\; \vvv_1^{d0,p}, \text{ for }\; z \ll d.
 \esubalat
Following the procedure of Ref.~\onlinecite{Bach2018}, we introduce the integrals $\intn ab$ of the integrand $a(z)\,b(z)^*$, where $a(z)$ and $b(z)$ are any of the functions $1$, $q(z)$, and $r(z)$,
 \bal
 \eqlab{def_intn}
 \intn ab &=  \int^{z}\!\dm z_n \int^{z_n}\!\dm z_{n-1} \ldots \int^{z_2}\!\dm z_1\:
 a(z_1)\: b(z_1)^* \bigg|^{{}}_{z=0},
 \nn
 \\
 \intn ab & \propto \delta^n,
 \text{ with }\, \delta = \dels,\delt \text{ and }
 n = 1, 2, 3, \ldots.
 \eal
With this notation, \eqref{JSB39a_leading} is easily integrated to give
 \bsub
 \bal
 \eqlab{v2thInterim}
 \vvv_{2,\parallel}^{\delta 0,T} &=
 \frac{\alfPO \DthO}{\nu_0 \delt^2}  \re\!\bigg[
 \frac{\delt + \ii\dels}{\dels}\,  \intII qr   \vvv_1^{\delta 0} T_1^{\delta 0*}
 + \ii \intII 1r   \vvv_1^{d0,p}T_1^{\delta 0*}\bigg]_\parallel
 \nn
 \\
 & \quad -\frac1{\etaO\dels^2} \re\!\bigg[
 \frac{\dels + \ii\delt}{\delt}\, \intII rq   \eta_1^{\delta 0} \vvv_1^{\delta 0*}
 + \ii \intII 1q   \eta_1^{d0} \vvv_1^{\delta 0*}\bigg]_\parallel.
 \eal
where the integrals are given by $I_{ba}^{(n)} = \big[I_{ab}^{(n)}\big]^*$ and
 \beq{IintVal}
 I_{1r}^{(2)} = -\frac{\ii}{2}\delt^2, \;\;
 I_{1q}^{(2)} = -\frac{\ii}{2}\dels^2, \;\;
 I_{rq}^{(2)} = \frac{\ii\,\dels^2\delt^2}{2(\dels+\ii\delt)^{2}}.
 \eeq
 \esub
When inserting $\rho_1^\delta = -\rhoO \alphap T_1^\delta$ in the final expression for the thermal correction, $\vvv_{2\parallel}^{\delta 0,T}$ becomes
\bal
 \eqlab{v2thFinal}
 \vvv_{2,\parallel}^{\delta 0,T}
 &= -\frac{1}{2\rho_0}
 \frac{\delt^2}{\dels^2} \re\Big[ \frac{\dels}{\dels-\ii\delt} \vvv_1^{\delta 0}\rho_1^{\delta 0*}
 + \vvv_1^{d0,p}\rho_1^{\delta 0*}\Big]_\parallel
 \nn
 \\
 & \quad  -\frac{1}{2 \eta_0} \re \Big[
 \frac{\delt}{\delt-\ii \dels}\,\eta_1^{\delta 0}  \vvv_1^{\delta 0*}
 + \eta_1^{d0} \vvv_1^{\delta 0*}\Big]_\parallel,
\eal
where two terms are due to the change in density and two to the change in viscosity.
The perpendicular part of the short-ranged streaming velocity $v_{2z}^{\delta 0,T}$ can be found by integrating the continuity equation~\eqnoref{2nd_order_ContEqu}, $\pp_z v_{2z}^{\delta,T} = - \nablabf_{\parallel}\cdot \vvv_{2\parallel}^{\delta,T} - \frac{1}{\rho_0}\nablabf \cdot \avr{\rho_1 \vvv_1}^{\delta,T}$, once with respect to $z$,
 \beq{JSB39b}
 v_{2z}^{\delta,T} = - \nablabf_{\parallel}\cdot \int^z \vvv_{2\parallel}^{\delta,T}\:\dm z - \frac{1}{\rho_0} \int^z \nablabf \cdot \avr{\rho_1 \vvv_1}^{\delta,T}\:\dm z.
 \eeq
The term $\int^z \vvv_{2\parallel}^\delta\:\dm z$ is given by \eqref{v2thInterim} by substituting all $\intII ab$ by $\intIII ab \propto \intII ab \delta $, so $\big|\nablabf_{\parallel}\cdot \int^z \vvv_{2\parallel}^\delta\:\dm z\big| \sim (k_c\delta)\,\big|\vvv_{2\parallel}^{\delta0}\big|$, and $\int^z \nablabf \cdot \avr{\rho_1 \vvv_1}^\delta\:\dm z \approx \int^z \pp_z\avr{\rho_1^\delta v_{1,z}^{d,p}}\:\dm z = \avr{\rho_1^{\delta} v_{1,z}^{d,p}}$. Including pre-factors, we obtain to leading order in $k_c\delta$,
\beq{v2del0T}
 v_{2,z}^{\delta 0,T} =- \frac{1}{2\rho_0}\re\!\Big[ \rho_1^{\delta 0*} v_{1z}^{d0,p}\Big].
 \eeq

\subsection{Bulk field and effective boundary condition}
With the short-range boundary-layer streaming term $\vvv_2^{\delta 0} = \vvv_2^{\delta 0,p} + \vvv_2^{\delta 0,T}$ in place, it is now possible to set up the governing equations and boundary conditions for the second-order bulk acoustic streaming $\vvv_2^d$,
 \bsubal{v2d_govA}
 \eqlab{v2d_cont_equ}
 0&=\nablabf \cdot \big( \rho_0 \vvv_2^d \big) - \rhoac^d,
 \\
 \eqlab{v2d_Stokes_equ}
 0&= - \nablabf p_2^d +\nablabf \cdot \taubf_2^d + \fachat^d,
 \\
 \eqlab{v2d_tau11}
 \taubf_2^d &= \eta_0 \big[ \nablabf \vvv_2^d + (\nablabf \vvv_2^d)^\dagger\big]
             + \beta \eta_0 (\nablabf \cdot \vvv_2)\, \III ,
 \\
 \eqlab{v2d0_bcA}
 \vvv_2^{d0} &= -\vvv_2^{\delta0}
 -\frac{1}{\omega}\avr{\big(\ii\VVV^0_1\cdot\nablabf\big)\vvv_1}\big|_{\rrr = \sss_0}.
 \esubal
Here, $\rhoac^d$ and $\fachat^d$ are the bulk terms in \eqref{acoustic_contribution}.
In the mass-conservation equation,  $\divop\vvv_2^d$  becomes,
 \beq{divv2d}
 \nablabf \cdot \vvv_2^d = -\frac{\nablabf \cdot \avr{\rho_1^{d} \vvv_1^{d,p}}}{\rho_0} = \Gamma \frac{k_0 \abs{\vvv_1^{d,p}}^2}{2 c_0}.
 \eeq
Each term of $\nablabf \cdot \vvv_2^d$ scales as $\frac{k_0}{c_0}  |\vvv_1^{d,p}|^2 \gg \frac12 \Gamma \frac{k_0}{c_0}\vert \vvv_1^{d,p}  \vert^2$, so $\frac{1}{\rhoO}\divop\avr{\rho_1^{d} \vvv_1^{d,p}}$ is negligible compared to the individual terms in $\nablabf \cdot \vvv_2^d$. We thus conclude that $\nablabf \cdot \vvv_2^d =0$, and that the streaming flow is incompressible. The acoustic body force $\fachat^d$ may be expressed as follows, where $\nablabf \rhoO$ and $\nablabf \kappa_{s0}$ unlike in previous work \citep{Karlsen2016, Karlsen2018} can be induced by temperature gradients,
 \bsubal{facd_in}
 \fachat^d &= -\nablabf \cdot \avr{\rho_0 \vvv_1^{d,p} \vvv_1^{d,p}} + \nablabf\cdot \taubf_{11}^d
 \\
 \eqlab{rhovv}
 &=-\nablabf \avr{\Lac^d} +\frac{1}{4}\abs{\vvv_1^{d,p}}^2\nablabf \rho_0 +\frac{1}{4}\abs{p_1}^2\nablabf \kapsO
 \nn\\
 &\qquad \qquad \qquad \quad
 -\frac{\Gamma \omega}{c_0^2} \avr{\vvv_1^d p_1} + \divop \taubf_{11}^d,
 \esubal
The gradient force $-\nablabf \avr{\Lac^d}$ of the Lagragian $\avr{\Lac^d}=\frac{1}{4} \kapsO |p_1|^2-\frac{1}{4} \rho_0 |\vvv_1^d|^2$ does not induce streaming. \citep{Riaud2017a, Bach2018} The next two terms form the inhomogeneous acoustic body force spawned by gradients in the density $\rho_0$ and in the compressibility $\kapsO$. \citep{Karlsen2016} The subsequent Eckart-streaming force term is important for either large systems or for rotating acoustic waves where $\vvv_1^d$ and $p_1$ have significant in-phase components. \citep{Bach2019} The last contribution $\divop \taubf_{11}^d$ is due to the temperature-dependent viscosity, $\eta_1^d = a_\eta \etaO \alfPO T_1^d = a_\eta (\gamma-1)\etaO\kapsO p_1$. Using $\vvv_1^{d,p} \approx -\ii\frac1{\omega \rhoO} \nablabf p_1$ as well as $\divop\big[\nablabf \vvv_1^{d,p} +(\nablabf \vvv_1^{d,p} )^\dagger \big] = 2\nablabf(\divop\vvv_1^{d,p}) =  -2k_c^2\vvv_1^{d,p}$, $\nablabf \eta_1^d = a_\eta \eta_0 (\gamma-1)\frac{\ii k_c}{\cO}\,\vvv_1^{d,p}$, and $\avr{\eta_1 (\divop\vvv_1^{d,p})} \propto \avr{p_1 (\ii p_1)} =0$, we reduce $\nablabf\cdot \taubf_{11}^d$ to
 \beq{nabsigma11}
 \nablabf\cdot \taubf_{11}^d
 =
 2(\gamma\!-\!1) a_\eta \eta_0  \frac{\omega^2}{\cOsqr}
 \Big[\avr{\big(\frac{\ii}{\omega}\vvv_1^{d,p}\!\cdot\!\nablabf\big) \vvv_1^{d,p}}\!
 -\!\kapsO \avr{\vvv_1^{d,p} p_1 }\Big].
 \eeq
Here, the first and second term involve the Stokes drift and the classical Eckart attenuation~\eqnoref{rhovv}, respectively. Now, collecting the results~\eqnoref{divv2d}-\eqnoref{nabsigma11}, the governing equations~\eqnoref{v2d_cont_equ}-\eqnoref{v2d_tau11} of the acoustic streaming become,
 \bsubal{v2d_gov}
 \eqlab{v2d}
 0&=\nablabf \cdot  \vvv_2^d ,
 \\
 \eqlab{p2d}
 0&=- \nablabf\big[p_2^d -\avr{\Lac^d} \big] + \eta_0 \nabla^2 \vvv_2^d +\fac^d,
 \\
 \eqlab{facd}
 \fac^d &=
 -\frac{1}{4}\abs{\vvv_1^{d,p}}^2\nablabf \rho_0 -\frac{1}{4}\abs{p_1}^2\nablabf\kapsO
 \nn
 \\
 & \quad +\left[1-\frac{2 a_\eta (\gamma-1)}{\beta+1} \right]
 \frac{\Gamma \omega}{\cOsqr} \avr{\vvv_1^{d,p} p_1}
 \nn
 \\
 &\quad + 2a_\eta \eta_0 (\gamma -1) \frac{\omega}{\cOsqr}
 \avr{\ii\vvv_1^{d,p} \cdot \nablabf \vvv_1^{d,p}}.
 \esubal
Here, the Lagrangian density $\avr{\Lac^d}$ is merged with  $p_2^d$ as an excess pressure. Since  $\nablabf\avr{\Lac^d}$ is orders of magnitude larger than $\fac^d$, its merging with $\nablabf p_2^d$ renders the numerical simulation more accurate,\citep{Riaud2017a} and makes it possible to use a coarser mesh in the bulk of the fluid domain.\citep{Bach2018} The term $-\frac{2 a_\eta (\gamma-1)}{\beta+1}\approx 0.44$ leads to an increase of the bulk-driven Eckart streaming by $44\%$ compared to a purely viscous model. The last term is due to gradients in the viscosity $\eta_1^d$, so a fluid particle oscillating $\sss_1=\frac{i}{\omega}\vvv_1^{d,p}$ experiences a varying viscosity during its oscillation period.

Finally, the thermal corrections to the boundary condition~\eqnoref{v2d0_bcA} stem from
$\vvv_2^{\delta 0,T}$ in the boundary-layer velocity $\vvv_2^{\delta 0} = \vvv_2^{\delta 0,p} + \vvv_2^{\delta 0,T}$, see \eqsref{v2thFinal}{v2del0T}, and from $\vvv_1^{d,T}$ in $\vvv_1 =  \vvv_1^\delta + \vvv_1^{d,p} + \vvv_1^{d,T}$ in the Stokes drift term
$-\frac{1}{\omega}\avr{\big(\ii\VVV^0_1\cdot\nablabf\big)\vvv_1}\big|_{\rrr = \sss_0}$. As $\big|\vvv_{1\parallel}^{d,T}\big| \ll \abs{v_{1z}^{d,T}}$, then $\VVV_1^0\cdot\nablabf\vvv_{1}^{d,T} \approx V_{1,z}^0 \alfPO \DthO \ppsqr_zT_1^\delta\,\een_z = \frac{\omega}{\rhoO}V_{1,z}^0\big(\ii\rho_1^\delta\big)\,\een_z$,
 \beq{ThermalStokesDrift}
 \frac{1}{\omega}\avr{\big(\ii\VVV_1^0\cdot\nablabf\big) \vvv_{1}^{d,T}}\Big|_{\rrr=\sss_0}
 = \frac{1}{2 \rho_0} \re\!\big[V_{1,z}^0 \rho_1^{\delta 0*}\big]\,\een_z.
 \eeq
In terms of the $\AAA$- and $\BBB$-vector notation of Ref.~\onlinecite{Bach2018}, the boundary condition~\eqnoref{v2d0_bcA} for the streaming velocity $\vvv_2^d$ is given by the purely viscous terms (superscript \qmarks{vs}) from Ref.~\onlinecite{Bach2018} and the thermal corrections (superscript \qmarks{th}) due to $\vvv_2^{\delta 0,T}$, \eqsref{v2thFinal}{v2del0T}, and
$\vvv_1^{d0,T}$, \eqref{ThermalStokesDrift},
 \bsubal{v2d_AB}
 \vvv_2^{d0} &=  \left(\AAA \cdot \eee_x \right)\eee_x + \left(\AAA \cdot \eee_y \right)\eee_y+ \left(\BBB \cdot \eee_z \right)\eee_z,
 \\
 \nn
 & \text{ with } \AAA = \AAA^\mr{vs}+\AAA^\mr{th}, \quad \BBB = \BBB^\mr{vs}+\BBB^\mr{th},
 \\
 \AAA^\mr{vs} &= - \frac{1}{2 \omega} \re \bigg[\! \vvv_1^{\delta 0 *} \cdot \nablabf \Big({\small \frac{1}{2}} \vvv_1^{\delta 0} -\ii \VVV_1^0 \Big) -\ii \VVV_1^{0*} \cdot \nablabf \vvv_1^{d,p}
 \nn
 \\
 &\quad+\!\bigg\{\! \frac{2\!-\!\ii}{2} \nablabf \!\cdot\! \vvv_1^{\delta 0 *}
 \!+ \ii \big( \nablabf \!\cdot\! \VVV_1^{0 *} \!\!- \pp_z v_{1z}^{d,p *}\big)\!\bigg\}
 \vvv_1^{\delta0}\!\bigg],\!\!
 \\
 \AAA^\mr{th} &= \frac{1}{2\rho_0} \frac{\delt^2}{\dels^2}
  \re \bigg[ \frac{\dels}{\dels-\ii\delt}  \vvv_1^{\delta 0}\rho_1^{\delta 0*}
  + \vvv_1^{d0,p}\rho_1^{\delta 0*}\bigg]
 \nn
 \\
 &\quad
  + \frac{1}{2 \eta_0} \re \bigg[\frac{\delt}{\delt-\ii \dels} \eta_1^{\delta 0}  \vvv_1^{\delta 0*}
  +\eta_1^{d0} \vvv_1^{\delta 0*}\bigg]
 \\
 \BBB^\mr{vs} &= \frac{1}{2 \omega}  \re \Big[ \ii \vvv_1^{d0,p *} \cdot \nablabf \vvv_1^{d,p} \Big]\\
 \BBB^\mr{th} &= \frac{1}{2\rho_0}\re\!\Big[\big(\vvv_{1}^{d0,p}-  \VVV_{1}^0\big) \rho_1^{\delta 0*}\Big].
 \esubal
The magnitude of the thermal terms are $(\gamma-1)a_q$ times the magnitude of the leading viscous terms. For water, $(\gamma-1)|a_\eta| \approx 0.9$ and $(\gamma-1)|a_\rho| \approx 0.01$ at room temperature, so here, the $\eta_1$-terms are important and must be included in acoustofluidic analyses, whereas $\rho_1$-terms are negligible. For gases with $\gamma-1 \approx 0.4$, the density terms may be important.

The results in \eqsref{v2d_gov}{v2d_AB} are our main results for the second-order streaming part of the effective thermoviscous theory, and they form the equations that are implemented in our numerical model.

\section{Numerical implementation and examples}
\seclab{examples}
We implement the effective thermoviscous model in the commercial finite-element software COMSOL Multiphysics. \citep{Comsol56} It is validated by comparisons to full numerical simulations, and two examples of significant thermal effects in acoustofluidic devices are shown. All simulations are done in COMSOL 5.6 \citep{Comsol56} on a HP-G4 workstation with a processor Intel Core i9-7960X @ $4.20\, \SIGHz$ and with 128 GB ram.

The effective thermoviscous model solver contains three steps: (1) the zeroth-order thermal field, (2) the acoustic pressure  and displacement fields, and (3) the stationary streaming fields. The acoustic temperature field $T_1$ is included analytically and therefore does not increase the numerical workload compared to the purely viscous model. The effective thermoviscous theory allows us to simulate acoustofluidic systems in 3D, which has prohibitive numerical costs for the full model.

Following our previous work,\citep{Muller2014, Karlsen2016, Ley2017, Bach2018, Skov2019} the governing equations
\eqnoref{energy0th}, \eqnoref{Helmholtz}, \eqnoref{Helmholtzsl}, and \eqnoref{v2d_gov} are implemented in COMSOL using the mathematical PDE module. The surface fields (superscript \qmarks{0}) are defined only on the fluid-solid interfaces. The effective boundary conditions \eqnoref{p1u1BCfinal} for $p_1$ and  $\uuu_1$ are implemented as weak contributions, whereas the boundary condition~\eqnoref{v2d_AB} for $\vvv_2^d$ is implemented as a Dirichlet boundary condition.

\begin{figure}[!t]
 \centering
 \includegraphics[width=\columnwidth]{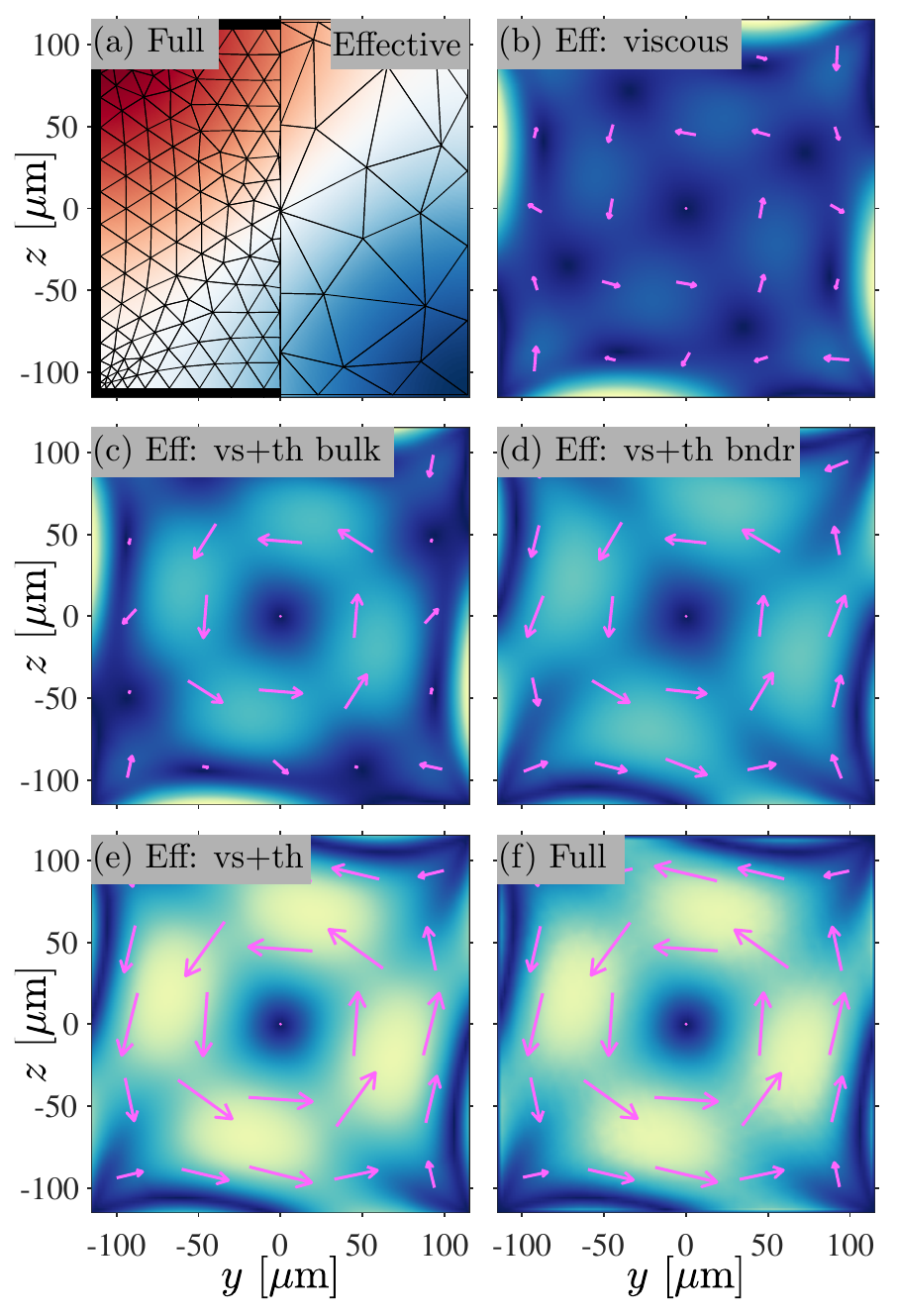}
 \caption[]{\figlab{square_fig}
 Simulated fields in a square channel with a rotating pressure wave of energy density $\Eac = 19\, \SIJ/\SIm^3$ actuated as described in the text.  (a) Color plot of $p_1$ at time $t=0$ from $-0.4$ (blue) to  $+0.4\,\SIMPa$ (red), and the mesh used in the full (left) and in the effective (right) thermoviscous model . (b) Vector plot of the streaming velocity $\vvv_2$ (magenta) and color plot of its magnitude from 0 (dark blue) to $20~\SImum/\SIs$ (yellow) [same scale in (b)-(f)] for the effective viscous model without thermal terms. (c) $\vvv_2$ for the effective viscous model with thermal bulk terms. (d) $\vvv_2$ for the effective viscous model with thermal boundary terms. (e) $\vvv_2$ for the complete effective thermoviscous model. (f) $\vvv_2$ for the full thermoviscous model.}
 \end{figure}

\subsection{Example I: 2D streaming in a square channel}
\seclab{ex1}

The first example is the square channel, which has been studied both experimentally \citep{Antfolk2014, Mishra2014, Gralinski2014} and numerically.\citep{Antfolk2014} In a square square channel, a rotating acoustic wave can be set up by two perpendicular, out-of-phase standing waves, as analyzed theoretically by Bach and Bruus. \citep{Bach2019} We apply the effective thermoviscous model in the fluid domain of the square channel in the 2D $yz$ cross section with the velocity $\VVV^0_1 = V_0\ee^{-\iot}\een_y$ at the vertical sides $y=\pm\frac12W$ and $\VVV^0_1 = \ii V_0\ee^{-\iot}\een_z$ at the horizontal sides $z=\pm\frac12 H$, a rigid-wall model with side length $H=W=230\,\SImum$. The zeroth-order temperature field is set to be constant, $T_0 = 20\, \SICel$.  We emphasize three main points of the results, shown in \figref{square_fig}: (1) The effective thermoviscous model reduces the computational time and memory requirements significantly. (2) Given that it is 2D, the full model can be simulated, and it agrees with and thus validates the effective model. (3) The thermal corrections strongly influence the streaming flow pattern.

The meshes plotted on top of the pressure field in \figref{square_fig}(a) are the ones needed to obtain an  $L_2$-norm-convergence\citep{Muller2014} of $0.1 \,\%$ for $p_1$ and $1\, \%$ for the streaming $\vvv_2$ for the full and for the effective model. With computation times of 15\,s versus 2\,s and 130042 degrees of freedoms versus 1788, the effective model is in this case 7 times faster and requires 130 times less memory than the full mode to achieve the same accuracy. \figref{square_fig}(b)-(f) show the resulting streaming $\vvv_2$ obtained using different assumptions. Panels (e) and (f) illustrate that the effective and full models agree, thus validating the former. Panel (b) shows how much $\vvv_2$ is changed when disregarding all thermal effects as in Ref.~\onlinecite{Bach2018}, whereas panel (c) and (d) illustrate the effect of adding only the thermal bulk effects of \eqref{v2d_gov}, and adding only the thermal correction to the boundary condition~\eqnoref{v2d_AB}.  Clearly, all the thermal effects need to be added, and in this example they stem from the temperature dependence of the viscosity through $\eta_1$ in the bulk term~\eqnoref{facd_in} $\nablabf \cdot \taubf_{11}$ and the boundary term~\eqnoref{v2d_AB} $\AAA^{T}$. Physically, the bulk term strengthens the central streaming roll, whereas the boundary term changes the morphology of the boundary streaming and additionally strengthens the central streaming roll.

\subsection{Example II: 3D streaming due to thermal fields}
\seclab{ex2}

The second example is the capillary glass tube widely used as a versatile acoustic trap in many experimental studies. \cite{Hammarstrom2012, Lei2013, Mishra2014, Gralinski2014} Inside the tube, in the region above the piezoelectric transducer, a characteristic streaming flow pattern containing four horizontal flow rolls is established. \citep{Hammarstrom2012} This pattern cannot be explained in numerical modeling \citep{Lei2011, Ley2017} in terms of boundary-driven streaming or classical bulk Eckart streaming, but here we argue, based on our thermoacoustic simulation results, that thermal effects are responsible for this streaming pattern. This result is important as the streaming pattern is used to lead nanoparticles into the central region, where they are trapped by larger seed particles.

\begin{figure}[!t]
 \centering
 \includegraphics[width=\columnwidth]{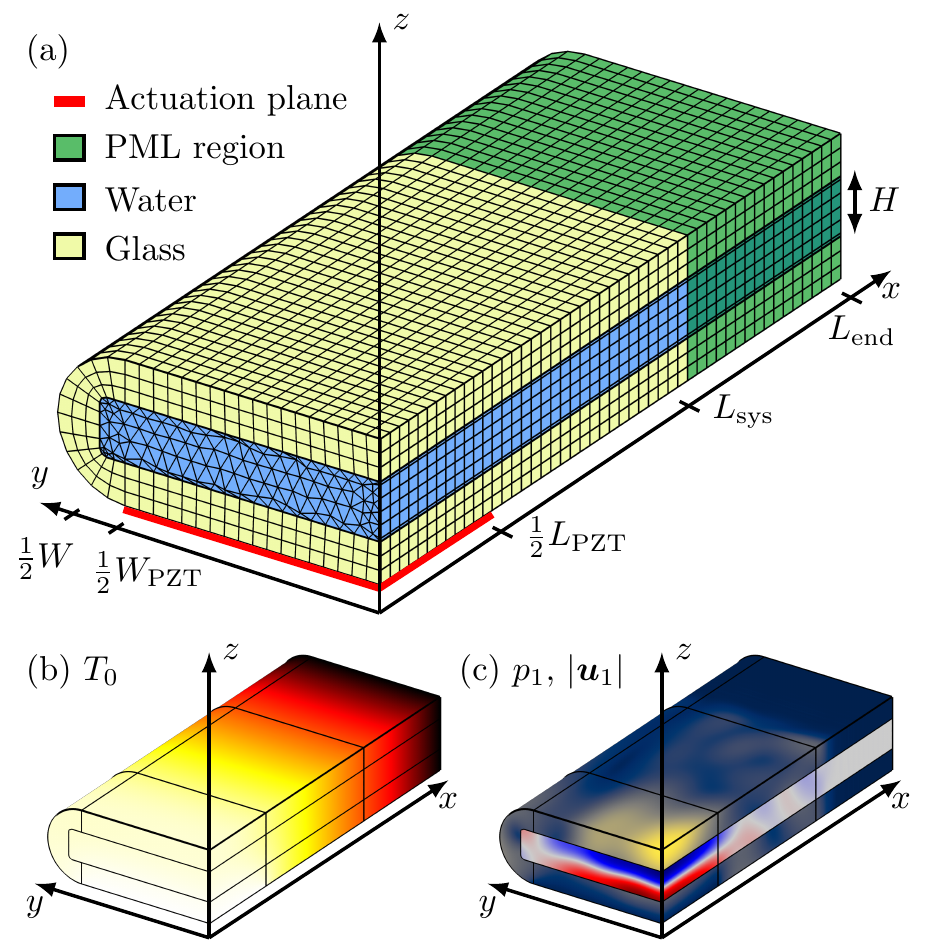} 
 \caption[]{\figlab{3D_system} (a) The simulated 3D system (reduced to a quarter by symmetry) consisting of the water (blue), the glass (yellow), and the artificially absorbing PML (green) domains. Also shown are the actuation region (red) and the mesh (black). (b) Color plot of the steady temperature $T_0$ from $20.0$ (black) to $21.5~\SICel$ (yellow). (c) Color plots of the displacement $\abs{\uuu_1}$ in the glass from 0 (blue) to 9~nm (yellow) and the acoustic pressure $p_1$ in the water from $-1.6$ (blue) to $+1.6~\SIMPa$ (red). Note the dampening of $\uuu_1$ and $p_1$ in the PML region.
 }
 \end{figure}

The 3D model, see \figref{3D_system}, is similar to device C1 in our previous work:\citep{Ley2017} a glass capillary tube of width $W=2$\,mm and height $H=0.2$\,mm, actuated from below in its central region by a piezoelectric transducer. The temperature is set to $T_\mr{air} = 20\,\SICel$ at $x=L_\mr{end}$ and to zero flux on all other outer surfaces except on the transducer. For simplicity, the transducer is represented by a (red) region of width $W_\mr{PZT}$, length $L_\mr{PZT}=1.16\, \SImm$ on the glass surface, with a given oscillatory displacement  $\uuu = \uuu_\mr{PZT}\ee^{-\iot}$ and steady temperature \citep{Werr2019} $T = T_\mr{air} + T_\mr{PZT}$, where $\uuu_\mr{PZT} = u_0\,\een_z$ with $u_0 = 0.25$\,nm and $T_\mr{PZT}=1.5\, \SICel$. We exploit the $xz$ and $yz$ symmetry planes and simulate only a quarter of the system. To simulate an infinitely long channel, we use a perfectly matched layer (PML) with artificial dampening to avoid reflections.\citep{Collino1998, Ley2017} The mesh shown in \figref{3D_system}(a) results in an $L_2$-norm-convergence\citep{Muller2014} of 1\,\% in the pressure $p_1$ and in the streaming $\vvv_2$, and of 3\,\% in the displacement $\uuu_1$. The simulation requires 491.959 degrees of freedom and takes 7~minutes.

For the steady temperature $T_0$ shown in \figref{3D_system}(b), we find by inspection a resonance at $f = 3.898\,\SIMHz$, for which the resulting acoustic displacement $\uuu_1$ and pressure $p_1$ are shown in \figref{3D_system}(c). $T_0$ is inhomogeneous with an almost constant temperature gradient along the tube in the $x$-direction, and, in agreement with previous experiments\citep{Hammarstrom2012} and simulations, \citep{Ley2017} $p_1$ appears as a vertical half-wave resonance localized in the region above the transducer, but stronger in the center than at the sides. Combining the effects of $p_1$ and the $T_0$-dependency of the density $\rho_0$ and compressibility $\kapsO$, the acoustic body force~\eqnoref{facd} driving the streaming $\vvv_2$ in the water domain becomes
 \bal
 \eqlab{fac_T0}
 \fac^{d} &\approx -\frac{1}{4}\abs{\vvv_1}^2\nablabf \rho_0 -\frac{1}{4} \abs{p_1}^2 \nablabf \kapsO
 \nn
 \\
 & = -\frac{1}{4} \Big(a_\rho \rho_0  \abs{\vvv_1}^2+ a_\kappa \kapsO   \abs{p_1}^2\Big) \alfPO  \nablabf T_0
 \eal
Since by \eqref{a_water}, $\kaps$ has a stronger temperature dependency than $\rho$, $\fac^{d}$ is dominated by the $\abs{p_1}^2$-term. This results in a body force parallel to $\grad T_0$ and strongest in the center,  where $\abs{p_1 }$ is maximum.

\begin{figure}[!t]
 \centering
 \includegraphics[width=\columnwidth]{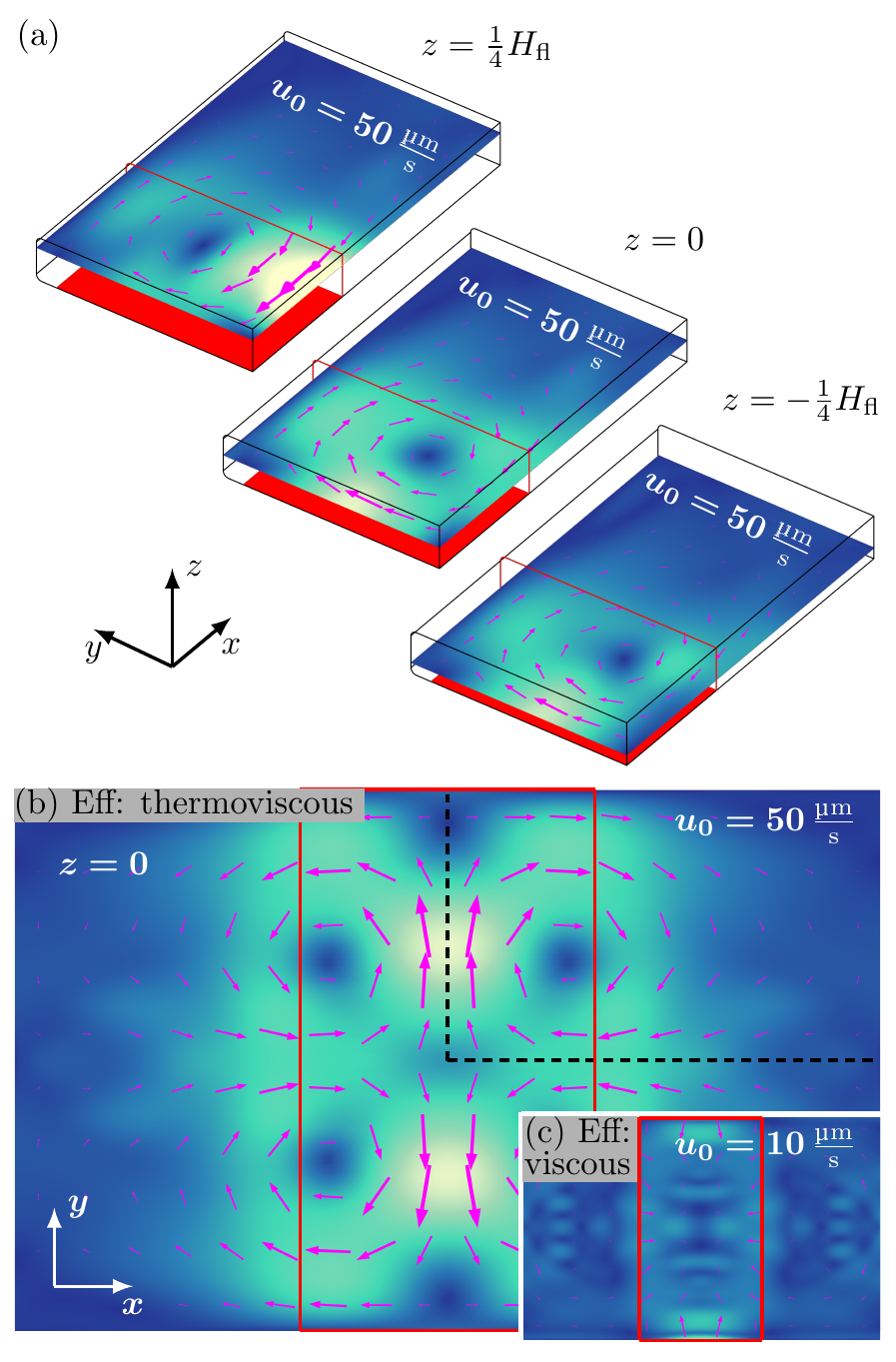} 
 \caption[]{\figlab{3D_streaming} The streaming velocity $\vvv_2$ (magenta arrows) and its magnitude from 0 (blue) to $ u_0 = 50\,\SImum/\SIs$ (yellow) in a symmetry quarter of the trapping capillary tube. (a) $\vvv_2$ in three different horizontal planes. (b) $\vvv_2$ in the full central plane $z=0$. The dashed black lines show the symmetry planes, and the red lines the edge of the actuation region. (c) $\vvv_2$ in the central plane $z=0$ without thermal effects. Note that here $u_0 = 10\,\SImum/\SIs$.
 }
 \end{figure}

The numerical simulation result for $\vvv_2$ is shown in \figref{3D_streaming}: The characteristic four horizontal flow rolls are clearly seen, the radius of which are determined by the width of the channel and the width of the actuation as observed by Hammarstr\"om \textit{et al.} \citep{Hammarstrom2012} This phenomenon is explained in terms of the acoustic body force $\fac^{d}$, which pushes the liquid into the center region near the vertical $xz$-plane at $y=0$, where it is strongest, accompanied by a back-flow at the edges near $y = \pm\frac12 W$, where the body force is weaker. In \figref{3D_streaming}(a) $\vvv_2$ is shown in three different horizontal planes. The variation in the flow rolls reflects the $z$-dependence of the thermal gradient above the transducer. In \figref{3D_streaming}(b), $\vvv_2$ is shown in the full horizontal plane at $z = 0\,\SImum$. Note, how the four flow roll centers are located near the edge (red lines) of the actuation region. To emphasize the crucial role of the thermal effects, we show in \figref{3D_streaming}(c) the streaming flow resulting from neglecting all thermal effects: in agreement with previous purely viscous models, but in contrast to experimental observations, the characteristic four-flow-roll pattern does not appear. Another important feature of the thermoviscous streaming is its magnitude. In \figref{3D_streaming}, $\abs{\vvv_2} = 50\,\SImum/\SIs$ is obtained with an acoustic energy density of $\Eac = 73\,\SIJ/\SIm^3$. This is five times larger than the  $10\,\SImum/\SIs$ of the purely viscous streaming, and notably only a factor of 3 lower than the 150-$\SImum/\SIs$-limit of \secref{validity_range} that marks the validity of the applied effective thermoviscous model.

In conclusion, the example highlights two important aspects: (1) The effective thermoviscous model enables 3D thermoviscous simulations in acoustofluidic systems, and (2) even moderate thermal gradients may create high streaming velocities in acoustofluidic systems. Such gradients can of course be created not only by heat generation in the transducer as in this example, but also more controllable by ohmic wires, Peltier elements, and external light sources. Notably, the validity of the perturbative approach breaks down at moderately high, but experimentally obtainable acoustic energy densities above $\sim 100~\SIJ/\SIm^3$ in combination with a moderate thermal gradient $\sim 1~\SIK/\SImm$, and this calls for an extension beyond perturbation theory of the presented theory.

\section{Conclusion}
\seclab{conclusion}

We have derived an effective thermoviscous theory for a fluid embedded in an elastic solid. The steady zeroth order temperature field is governed by \eqref{energy0th}. The acoustic fields are governed by the Helmholtz equations \eqsnoref{Helmholtz}{Helmholtzsl}, the decompositions \eqsnoref{v1T1}{u1T1}, and the effective boundary conditions \eqnoref{p1u1BCfinal}. The time-averaged acoustic streaming is governed by the effective Stokes equation~\eqnoref{v2d_gov} and the effective boundary conditions~\eqnoref{v2d_AB}. The theory includes the thermoviscous boundary layers and the acoustic temperature field $T_1$ analytically, and impose them as effective boundary conditions and time-averaged body forces on the thermoacoustic bulk fields.

The theory has been implemented in a numerical model, which because it avoids resolving numerically the boundary layers, allows for simulating both the first-order thermoviscous acoustic fields and second-order steady fields in 3D models of acoustofluidic systems. A conventional brute-force direct numerical simulations is very difficult, due to large memory requirements. In 2D, the model was validated by direct numerical simulations, and in 3D its self-consistency have been checked by mesh-convergence analyses.

We have applied the effective thermoviscous model in two numerical examples to demonstrate the importance of thermovisocus effects in microscale acoustofluidic devices. In particular, we have shown how the acoustic streaming depends strongly on the thermal fields: (1)~The oscillating temperature field $T_1$ impacts the streaming through the temperature dependency of the viscosity, causes corrections to the effective boundary condition, and spawns an additional body force in the bulk. In the 2D model of the square channel in \secref{ex1} and \figref{square_fig}, we have shown, how the thermoviscous effects are particularly important for the morphology and magnitude of the streaming in a rotating acoustic field. (2) The presence of an inhomogeneous stationary temperature field $T_0$ affects the streaming through the induced gradients in compressibility and density. In the 3D model of the capillary glass tube in \secref{ex2} and \figref{3D_streaming}, we have shown, how the experimentally-observed characteristic horizontal streaming rolls in the standing acoustic resonance of \figref{3D_system}, are caused by heating from the actuation area. We have also shown, how very high streaming velocities ($\sim 1\,\SImm/\SIs$) can be caused by small temperature gradients ($\sim 1\, \SIK/\SImm$) for moderate acoustic energy densities ($\sim 100\,\SIJ/\SIm^3$).

Our theoretical model enables 3D simulations of thermoviscous effects in microscale acoustofluidic devices. The results point to new ways for microscale hand\-ling of fluids and particles using a combination of acoustic and thermal fields. Although we have developed the effective thermoviscous theory within the narrow scope of microscale acoustofluidics, it is more general and may find wider use in other branches of thermoacoustics.

\acknowledgements

This work was supported by Independent Research Fund Denmark, Natural Sciences (Grant No.~8021-00310B).

%
%


\end{document}